\documentclass[11pt,twoside,nohyper]{pallis}

\usepackage{rotating}

\usepackage{epsfig}
\usepackage{latexsym}
\usepackage{amssymb}
\usepackage{fancyheda}
\usepackage{times}
\usepackage{caption}

\usepackage{slashed,cite}
\usepackage{upgreek}

\usepackage{amsmath}
\usepackage{amsfonts}
\usepackage{amsbsy}
\usepackage{amscd}
\usepackage{bbm}

\newcommand{\bal}{\begin{align}}
\newcommand{\eal}{\end{align}}
\newcommand{\beqs}{\begin{subequations}}
\newcommand{\eeqs}{\end{subequations}}
\newcommand{\eec}{\end{center}}
\newcommand{\bec}{\begin{center}}

\newcommand{\eem}{\end{matrix}}
\newcommand{\bem}{\begin{matrix}}
\newcommand{\eeq}{\end{equation}}
\newcommand{\beq}{\begin{equation}}
\newcommand{\ba}{\begin{array}}
\newcommand{\ea}{\end{array}}
\newcommand{\bea}{\begin{eqnarray}}
\newcommand{\eea}{\end{eqnarray}}
\newcommand{\baq}{\begin{eqnarray}}
\newcommand{\eaq}{\end{eqnarray}}

\newcommand{\Eref}[1]{Eq.~(\ref{#1})}
\newcommand{\Sref}[1]{Sec.~\ref{#1}}
\newcommand{\Srefs}[1]{Secs.~\ref{#1}}
\newcommand{\Fref}[1]{Fig.~\ref{#1}}
\newcommand{\Tref}[1]{Table~\ref{#1}}
\newcommand{\cref}[1]{Ref.~\cite{#1}}
\newcommand{\crefs}[1]{Refs.~\cite{#1}}

\newcommand\eqs[2]{Eqs.~(\ref{#1}) and (\ref{#2})}

\newcommand{\ftn}{\footnotesize}

\newcommand{\ssz}{\scriptsize}

\newcommand{\GeV}{{\mbox{\rm GeV}}}

\newcommand{\sFref}[2]{Fig.~\ref{#1}-{\ftn\sf ({#2})}}
\newcommand{\sEref}[2]{Eq.~(\ref{#1}{\ftn\sf {#2}})}

\newcommand{\etal}{{\it et al.\/}}

\def\to{\rightarrow}

\def\llgm{\left\lgroup}
\def\rrgm{\right\rgroup}
\def\lf{\left(}
\def\rg{\right)}
\newcommand\vev[1]{\langle {#1} \rangle}

\newcommand{\Vhi}{\ensuremath{\widehat V_{\rm HI}}}

\newcommand{\Hhi}{\ensuremath{\widehat H_{\rm HI}}}

\newcommand{\Khi}{\ensuremath{K}}

\newcommand{\Vhio}{\ensuremath{\widehat V_{\rm HI0}}}
\newcommand{\Ns}{\ensuremath{{\what N_\star}}}
\newcommand{\ck}{\ensuremath{c_{\rm K}}}

\newcommand{\mP}{\ensuremath{m_{\rm P}}}
\newcommand{\Mpq}{\ensuremath{M}}

\newcommand{\Mgut}{\ensuremath{M_{\rm GUT}}}
\newcommand{\Qef}{\ensuremath{\Lambda_{\rm UV}}}
\newcommand{\Ggut}{\ensuremath{G_{\rm GUT}}}

\def\openone{\leavevmode\hbox{\small1\kern-3.8pt\normalsize1}}
\newcommand{\dV}{\ensuremath{\Delta\widehat V_{\rm HI}}}

\newcommand{\fr}{\ensuremath{f_{\cal R}}}
\newcommand{\frs}{\ensuremath{f_{n\star}}}
\newcommand{\fms}{\ensuremath{f_{m\star}}}
\newcommand{\fns}{\ensuremath{f_{n\star}}}

\newcommand{\fk}{\ensuremath{f_{\rm K}}}
\newcommand{\fm}{\ensuremath{F_{-}}}

\newcommand{\ksp}{\ensuremath{k_{S-}}}
\newcommand{\fp}{\ensuremath{F_{+}}}
\newcommand{\fa}{\ensuremath{F_{1S}}}
\newcommand{\fb}{\ensuremath{F_{2S}}}
\newcommand{\ca}{\ensuremath{c_{\cal R}}}

\newcommand{\kx}{\ensuremath{k_S}}
\newcommand{\kpp}{\ensuremath{k_\Phi}}
\newcommand{\Dex}{\ensuremath{\Delta_{\rm max\star}}}

\newcommand{\Gsm}{\ensuremath{G_{\rm SM}}}

\newcommand{\msn}{\ensuremath{\what m_{\rm \dph}}}

\newcommand{\ks}{\ensuremath{k_\star}}
\newcommand{\ns}{\ensuremath{n_{\rm s}}}

\newcommand{\nb}{\ensuremath{N_{S}}}

\newcommand{\as}{\ensuremath{a_{\rm s}}}
\newcommand{\As}{\ensuremath{A_{\rm s}}}
\newcommand{\rw}{\ensuremath{r_{0.002}}}
\newcommand{\rs}{\ensuremath{r_{\pm}}}
\newcommand{\rpm}{\ensuremath{r_{\pm}}}

\newcommand{\rce}{\ensuremath{\widehat{\mathcal{R}}}}
\newcommand{\Ve}{\ensuremath{\widehat{V}}}

\newcommand{\dphi}{\ensuremath{\what{\delta\phi}}}
\newcommand{\dph}{\ensuremath{\delta\phi}}

\newcommand{\phc}{\ensuremath{\Phi}}
\newcommand{\phcb}{\ensuremath{\bar\Phi}}

\newcommand{\what}{\ensuremath{\widehat}}

\def\bbet{{\bar\beta}}
\def\al{{\alpha}}

\def\bt{{\beta}}

\def\th{{\theta}}
\def\thb{{\bar\theta}}

\def\thn{{\theta_{\Phi}}}

\newcommand{\Trh}{\ensuremath{T_{\rm rh}}}
\newcommand{\sg}{\ensuremath{\phi}}

\newcommand{\sgx}{\ensuremath{\phi_\star}}
\newcommand{\sgf}{\ensuremath{\phi_{\rm f}}}

\newcommand{\ld}{\ensuremath{\lambda}}
\newcommand{\ldu}{\ensuremath{\uplambda}}
\newcommand{\Ld}{\ensuremath{\Lambda}}
\newcommand{\kp}{\ensuremath{\kappa}}

\newcommand{\se}{\ensuremath{\widehat \phi}}
\newcommand{\sex}{\ensuremath{\widehat{\phi}_\star}}
\newcommand{\sef}{\ensuremath{\widehat{\phi}_{\rm f}}}
\newcommand{\geu}{\ensuremath{\widehat g}}
\newcommand{\eph}{\ensuremath{\widehat \epsilon}}
\newcommand{\ith}{\ensuremath{\widehat \eta}}

\newcommand\mtta[4]{\mbox{
$\llgm\bem #1 &#2 \cr #3& #4\eem\rrgm$}}

\def\trns{transplanckian}
\def\Kap{K\"{a}hler potential}
\def\Km{K\"{a}hler manifold}
\def\Kaa{K\"{a}hler~}

\def\str{Starobinsky}
\def\FHI{nMHI~}
\def\bcp{{\sc\small Bicep2}/{\it Keck Array}}
\newcommand{\plk}{{\it Planck}}

\newcommand{\diag}{\ensuremath{{\sf diag}}}

\newcommand{\cm}{\ensuremath{c_{-}}}
\newcommand{\cp}{\ensuremath{c_{+}}}

\newcommand{\Na}{\ensuremath{{N_1}}}
\newcommand{\Nb}{\ensuremath{{N_2}}}

\def\prdn#1#2#3#4{{\sl Phys. Rev. D }{\bf #1}, no. #4, #3 (#2)}

\renewenvironment{subequations}{%
\refstepcounter{equation}%
\setcounter{parentequation}{\value{equation}}%
  \setcounter{equation}{0}
  \ignorespaces
}{%
  \setcounter{equation}{\value{parentequation}}%
  \ignorespacesafterend
}

\title{\LARGE \bfseries\scshape Variants of Kinetically Modified Non-Minimal
Higgs Inflation in Supergravity}

\author{{\large \bfseries\scshape C. Pallis}\\
Department of Physics, University of Cyprus, \\ P.O. Box 20537,
Nicosia 1678, CYPRUS \\  {\sl e-mail address: }{\ftn\tt
cpallis@ucy.ac.cy}}

\abstract{We consider models of chaotic inflation driven by the
real parts of a conjugate pair of Higgs superfields involved in
the spontaneous breaking of a grand unification symmetry at a
scale assuming its \emph{Supersymmetric} (SUSY) value. Employing
\Kap s with a prominent shift-symmetric part proportional to $\cm$
and a tiny violation, proportional to $\cp$, included in a
logarithm we show that the inflationary observables provide an
excellent match to the recent \plk\ and \bcp\ results setting,
e.g., $6.4\cdot10^{-3}\lesssim\rs=\cp/\cm\lesssim1/N$ where $N=2$
or $3$ is the prefactor of the logarithm. Deviations of these
prefactors from their integer values above are also explored and a
region where hilltop inflation occurs is localized. Moreover, we
analyze two distinct possible stabilization mechanisms for the
non-inflaton accompanying superfield, one tied to higher order
terms and one with just quadratic terms within the argument of a
logarithm with positive prefactor $\nb<6$. In all cases, inflation
can be attained for subplanckian inflaton values with the
corresponding effective theories retaining the perturbative
unitarity up to the Planck scale.
\\ \\
{\ftn \sf Keywords: Cosmology of Theories Beyond the Standard
Model, Supergravity Models};\\
{\ftn \sf PACS codes: 98.80.Cq, 11.30.Qc, 12.60.Jv, 04.65.+e}
\\\\{\sl\bfseries Published in} {\sl J. Cosmol. Astropart. Phys. }{\bf 10}, 037,
no. 10, (2016)}

\begin{document}

\pagestyle{fancyplain}

\rhead[\fancyplain{}{ \bf \thepage}]{\fancyplain{}{\sl Variants of
Kinetically Modified nMHI}} \lhead[\fancyplain{}{\sl C.
Pallis}]{\fancyplain{}{\bf \thepage}} \cfoot{}

\section{Introduction}\label{intro}

In a series of recent papers \cite{nMkin, lazarides, nMHkin} we
established a novel type of \emph{non-minimal inflation} ({\sf\ftn
nMI}) called \emph{kinetically modified}. This term is coined in
\cref{nMkin} due to the fact that, in the non-SUSY set-up, this
inflationary model, based on the $\phi^p$ power-law potential,
employs not only a suitably selected coupling to gravity
$\fr=1+\ca\phi^{p/2}$ but also a kinetic mixing of the form
$\fk=\ck\fr^m$. The merits of this construction compared to the
original (and more economical) model \cite{old, sm1,nmi,atroest}
of nMI (defined for $\fk=1$) are basically two:

\begin{itemize}

\item[{\sf (i)}] The observational outputs depend on the ratio
$r_{{\cal R}K}=\ca/\ck^{p/4}$ and can be done excellently
consistent with the recent \plk\ \cite{plin} and \bcp\
\cite{gws,gwsnew} results;

\item[{\sf (ii)}] The resulting theory respects the perturbative
unitarity \cite{cutoff, riotto} up to the Planck scale for any $p$
and $m$ and $r_{{\cal R}K}\leq1$.

\end{itemize}

In the SUSY -- which means \emph{Supergravity} ({\sf\ftn SUGRA})
-- framework the two ingredients necessary to achieve this kind of
nMI, i.e., the non-minimal kinetic mixing and coupling to gravity,
originate from the same function, the \Kap, and the set-up becomes
much more attractive. Particularly intriguing is the version of
these models, called henceforth \emph{non-minimal Higgs inflation}
({\sf\ftn nMHI}), in which the inflaton, at the end of its
inflationary evolution, can also play the role of a Higgs field
\cite{old,jones2,nmH,okada,lazarides,nMHkin}. Actually in
\cref{nMHkin} we present a class of \Kap s which cooperate with
the simplest superpotential \cite{fhi} widely used for
implementing spontaneously breaking of a \emph{Grand Unified
Theory} ({\sf \ftn GUT}) gauge group $\Ggut$. In this framework,
the non-minimal kinetic mixing and gravitational coupling of the
inflaton can be elegantly realized introducing an approximate
shift symmetry \cite{shift, shiftHI, lazarides, nMHkin} respected
by the \Kap. As a consequence, the constants $\ck$ and $\ca$
introduced above can be interpreted as the coefficients of the
principal shift-symmetric term ($\cm$) and its violation ($\cp$)
while $r_{{\cal R}K}$ is now written as $\rs=\cp/\cm$ -- obviously
$p=4$ in this set-up.

Trying to highlight the most important issues of our suggestion in
\cref{nMHkin}, we employ only integer coefficients of the
logarithms appearing in the \Kap s. However, as we show in the
particular case of \cref{lazarides} -- see also \crefs{nIG, roest,
quad} -- the variation of the prefactors of the logarithms in the
\Kap s can have a pronounced impact on the inflationary
observables. Consequently, it would be interesting to investigate
which inflationary solutions can be obtained in this variance of
the simplified initial set-up. Moreover, we have here the
opportunity to test our proposal against the latest obervational
data on the gravitational waves \cite{gwsnew}. We also check the
wide applicability of a novel stabilization mechanism for the
non-inflaton accompanied field, recently proposed in the context
of the \str-type inflation in \cref{su11}. In addition to the
results similar to those found in \crefs{lazarides, nMHkin}, we
here establish a sizable region of the parameter space where nMHI
of hilltop type \cite{lofti} is achieved.

The super- and  \Kap s of our models are presented in
Sec.~\ref{fhim}. In Sec.~\ref{hi} we describe our inflationary
set-up, whereas in \Sref{fhi} we derive the inflationary
observables and confront them with observations. Our conclusions
are summarized in Sec.~\ref{con}. Throughout the text, we use
units where the reduced Planck scale $\mP = 2.433\cdot
10^{18}~\GeV$ is set to be unity, the subscript of type $,\chi$
denotes derivation \emph{with respect to} ({\small\sf w.r.t}) the
field $\chi$ -- e.g., $F_{,\chi\chi}=\partial^2F/\partial\chi^2$
-- and charge conjugation is denoted by a star ($^*$).

\section{Supergravity Set-up}\label{fhim}

The \emph{Einstein frame} ({\sf\ftn EF}) action within SUGRA for
the complex scalar fields $z^\al=S,\phc,\phcb$ -- denoted by the
same superfield symbol -- can be written as \cite{linde1}
\beqs \beq\label{Saction1} {\sf S}=\int d^4x \sqrt{-\what{
\mathfrak{g}}}\lf-\frac{1}{2}\rce +K_{\al\bbet} \geu^{\mu\nu}D_\mu
z^\al D_\nu z^{*\bbet}-\Ve\rg \eeq
where summation is taken over $z^\al$; $D_\mu$ is the gauge
covariant derivative, $K$ is the \Kap, with
$K_{\al\bbet}=K_{,z^\al z^{*\bbet}}$ and
$K^{\al\bbet}K_{\bbet\gamma}=\delta^\al_{\gamma}$. Also $\Ve$ is
the EF SUGRA potential which can be found via the formula
\beq \Ve=e^{\Khi}\left(K^{\al\bbet}D_\al WD^*_\bbet W^*-3{\vert
W\vert^2}\right)+\frac{g^2}2 \mbox{$\sum_a$} {\rm D}_a {\rm
D}_a,\label{Vsugra} \eeq \eeqs
where $D_\al W=W_{,z^\al} +K_{,z^\al}W$, with $W$ being the
superpotential, ${\rm D}_a=z_\al\lf T_a\rg^\al_\bt K^\bt$, $g$ is
the unified gauge coupling constant and the summation is applied
over the generators $T_a$ of $\Ggut$. Just for definiteness we
restrict ourselves to $\Ggut=G_{\rm SM}\times U(1)_{B-L}$
\cite{lazarides, nMHkin}, gauge group which consists the simplest
GUT beyond the \emph{Minimal SUSY Standard Model} ({\sf\ftn MSSM})
based on the gauge group ${G_{\rm SM}}= SU(3)_{\rm C}\times
SU(2)_{\rm L}\times U(1)_{Y}$ -- here ${G_{\rm SM}}= SU(3)_{\rm c}
\times SU(2)_{\rm L}\times U(1)_{Y}$ is the gauge group of the
standard model and $B$ and $L$ denote the baryon and lepton
number, respectively.

As shown in \Eref{Vsugra}, the derivation of $\Ve$ requires the
specification of $W$ and $K$ presented in \Srefs{fhim1} and
\ref{fhik} respectively. In \Sref{fhim3} we derive the SUSY
vacuum.

\subsection{Superpotential}\label{fhim1}

We focus on the simplest $W$ which can be used to implement the
Higgs mechanism in a SUSY framework. This is
\beq W=\ld S\lf\bar\Phi\Phi-\Mpq^2/4\rg\label{Win} \eeq
and is uniquely determined, at renormalizable level, by a
convenient \cite{fhi} continuous $R$ symmetry. Here $\ld$ and $M$
are two constants which can both be taken positive; $S$ is a
left-handed superfield, singlet under $\Ggut$; $\Phi$ and
$\bar\Phi$ is a pair of left-handed superfields which carry $B-L$
charges $1$ and $-1$ respectively and lead to a breaking of
$\Ggut$ down to $\Gsm$ by their \emph{vacuum expectation values}
({\sf\ftn v.e.vs}).

$W$ in \Eref{Win}, combined with a canonical \cite{fhi1} or
quasi-canonical $K$ \cite{fhi2, fhi3}, can support \emph{F-term
hybrid inflation} driven by $S$ with the $\phcb-\phc$ system being
stabilized at zero. This type of inflation is terminated by a
destabilization of the the $\phcb-\phc$ system which is led to the
SUSY vacuum during the so-called waterfall regime. Therefore, a
GUT phase transition takes place at the end of inflation.
Topological defects (cosmic strings in the case of $\Ggut$
considered here) are, thus, copiously formed if they are predicted
by the symmetry breaking. In our proposal we interchange the roles
of the inflaton and the waterfall fields attaining inflation
driven by $\phcb-\phc$ system and setting $S$ stabilized at the
origin during and after nMHI. As a consequence $\Ggut$ is already
spontaneously broken during nMHI through the non-zero values
acquired by $\phcb-\phc$ and so, nMHI is not followed by the
production of cosmic defects. To implement such an inflationary
scenario we have to adopt logarithmic $K$'s presented below.

\subsection{\Kaa\  Potentials}\label{fhik}

The implementation of the standard (large-field) nMHI
\cite{linde1, nmH} -- for small-field nMHI see \cref{okada} --
requires the adoption of a logarithmic $K$ including an
holomorphic term $\ca\phc\phcb$ in its argument together with the
usual kinetic terms. The resulting model has three shortcomings:
{\sf\ftn (i)} For $\ca\gg1$, the perturbative unitarity is
violated below $\mP$ \cite{cutoff,riotto}; {\sf\ftn (ii)} The
predicted $r$ lies marginally within the $1-\sigma$ region of
\bcp\ results \cite{gwsnew}; {\sf\ftn (iii)} Possible inclusion of
higher order terms of the form $|S|^2\lf k_{S\Phi} |\phc|^2+
k_{S\bar\Phi}|\phcb|^2\rg$ in $K$ generally violate \cite{talk}
the D-flatness unless an ugly tuning is imposed with
$k_{S\Phi}=k_{S\bar\Phi}$.

All the issues above can be overcome, as we show below, if we
assume the existence of an approximate shift symmetry on the $K$'s
along the lines of \cref{lazarides,nMHkin} -- the importance of
the shift symmetry in taming the so-called $\eta$-problem of
inflation in SUGRA is first recognized for gauge singlets in
\cref{shift} and non-singlets in \cref{shiftHI}. More
specifically, to achieve kinetically modified nMHI we select
purely or partially logarithmic $K$'s including the real functions
\beq \label{fs}
F_\pm=\left|\Phi\pm\bar\Phi^*\right|^2~~\mbox{and}~~\fa=|S|^2-\kx{|S|^4}
~~\mbox{or}~~\fb=1+|S|^2/\nb\,,~~\eeq
where, as we show in \Sref{hi1}, $\fm$ and $\fp$ are related to
the canonical normalization of inflaton and the non-minimal
inflaton-curvature coupling respectively. Also $\fa$ or $\fb$
provides typical kinetic terms for $S$, considering the
next-to-minimal term in $\fa$ for stability reasons \cite{linde1}.
In terms of the functions introduced in \Eref{fs} we postulate the
following form of $K$
\beqs\beq
K_1=-\Na\ln\left(1+\cp\fp-\frac1\Na(1+\cp\fp)^m\cm\fm-\frac1\Na
\fa+\kpp\fm^2+
\frac1\Na\ksp\fm{|S|^2}\right)\,.~~~~~\label{K1}\eeq
Here all the allowed terms up to fourth order are considered for
$\cp=0$. Switching on $\cp$ generates a violation of an enhanced
symmetry -- see below -- and gives rise to the scenario of
\emph{kinetically modified} nMHI as defined in \Sref{intro}.
Namely, the term $1+\cp\fp$ plays the role of the non-minimal
gravitational coupling whereas the factor $\cm(1+\cp\fp)^m$
dominates the nonminimal kinetic mixing. Other allowed terms such
as $\fp^m$ or $\fp^m |S|^2/3$ are neglected for simplicity or we
have to assume that their coefficients are negligibly small.
Identical results can be achieved if we place the first $\fm$ term
outside the argument of the logarithm selecting $K=K_2$ with
\beq K_2=-N_2\ln\left(1+\cp\fp-
\fa/\Nb\right)+(1+\cp\fp)^{m-1}\cm\fm\,.\label{K2}\eeq
If we place $\fa$ outside the argument of the logarithm in the two
$K$'s above, we can obtain two other $K$'s which lead to similar
results. Namely,
\bea K_3&=&-N_3\ln\lf1+\cp\fp-
(1+\cp\fp)^m\cm\fm/N_3\rg+\fa\,,\label{K3}\\
K_4&=&-N_4\ln(1+\cp\fp)+(1+\cp\fp)^{m-1}\cm\fm+\fa\,.\label{K4}\eea
If we employ $\fb$, the available $K$'s which lead to the same
outputs with the previous ones have the form of $K_3$ and $K_4$
replacing $\fa$ with $\nb\ln\fb$, i.e.,
\bea K_5&=&-N_5\ln(1+\cp\fp-(1+\cp\fp)^m\cm\fm/N_5)+\nb\ln\fb\,,\label{K5}~~~\\
K_6&=&-N_6\ln\left(1+\cp\fp\right)+(1+\cp\fp)^{m-1}\cm\fm\
+\nb\ln\fb\,.\label{K6}\eea
Furthermore, allowing the term including $\fm$ to share the same
logarithmic argument with $\fb$ we can obtain a last expression of
$K$, i.e.,
\bea
K_7=-N_7\ln(1+\cp\fp)+\nb\ln\lf\fb+(1+\cp\fp)^{m-1}\cm\fm/\nb\,\rg\,.\label{K7}
\eea \eeqs
The last three $K$'s are advantageous compared to the others since
the stabilization of $S$ is achieved with just quadratic terms and
so no higher order mix terms between $F_\pm$ and $S$ are necessary
for consistency.

As we show in \Sref{hi1}, the positivity of the kinetic energy of
the inflaton sector requires $\cp<\cm$ and $N_i>0$ with
$i=1,...,7$. For $\rs=\cp/\cm\ll1$, our models are completely
natural in the 't Hooft sense because, in the limits $\cp\to0$ and
$\ld\to0$, $K_i$ with $i=1,...,4$  enjoy the following enhanced
symmetries:
\beqs\beq \Phi \to\ \Phi+C,\>\>\>\bar\Phi \to\ \bar\Phi+C^*
\>\>\>\mbox{and}\>\>\> S \to\ e^{i\varphi} S,\label{shift}\eeq
where $C$ [$\varphi$] is a complex [real] number. In the same
limit, $K_i$ with $i=5$ and $6$ enjoy even more interesting
enhanced symmetries:
\beq \Phi \to\ \Phi+C,\>\>\>\bar\Phi \to\ \bar\Phi+C^*
\>\>\>\mbox{and}\>\>\>\frac{S}{\sqrt{\nb}}\to \frac{a
S/\sqrt{\nb}+b}{-b^*S/\sqrt{\nb}+a^*},\label{su2}\eeq
with $|a|^2+|b|^2=1$. In other words, for $K=K_5$ or $K_6$ the
theory exhibits a $SU(2)_S/U(1)$ enhanced symmetry. Besides this
symmetry, in the same limit, $K_7$ remains invariant (up to a \Kaa
transformation) under the continuous (non-holomorphic)
transformations
\beq  \frac{S}{\sqrt{\nb}}\to \frac{a
S/\sqrt{\nb}+b}{-b^*S/\sqrt{\nb}+a^*},\>\>\>\Phi \to\
\frac{\Phi}{-b^*S/\sqrt{\nb}+a^*}\>\>\>\mbox{and}\>\>\>\bar\Phi
\to\ \frac{\bar\Phi}{-b
S^*/\sqrt{\nb}+a}\cdot\label{su2new}\eeq\eeqs
The kinetic terms, though, do not respect this symmetry and so,
this is not valid at the level of the lagrangian.

In \Sref{res2} we scan numerically the full parameter space of the
models letting $m$ vary in the range $0\leq m\leq10$ and allowing
for a continuous variation of the $N_i$'s. On the other hand, we
have to remark that for $m=0$ [$m=1$], $\fm$ and $\fp$ in $K_1,
K_3$ and $K_5$ [$K_2, K_4, K_6$ and $K_7$] are totally decoupled,
i.e. no higher order term is needed. Given that the $m=0$ case
with $N_1\leq3$ or $N_3\leq2$ is extensively analyzed in
\cref{lazarides} we here focus mainly on $m=1$ with variable
$N_i$'s -- see \Srefs{res1b} and \ref{res2}. Moreover, keeping in
mind that the most well-motivated $K$'s from the point of view of
string theory are those with integer $N_i$'s -- cf. \cref{lust} --
we pay also special attention to the case with $N_i=3$ for $i=1,2$
or $N_i=2$ for $i=3,...,7$ -- see \Srefs{res1a} and \ref{res2}.

\subsection{SUSY Vacuum}\label{fhim3}

To verify that the theories constructed  lead to the breaking of
$G_{B-L}$ down to $\Gsm$, we have to specify the SUSY limit
$V_{\rm SUSY}$ of $\Vhi$ and minimize it. The potential $V_{\rm
SUSY}$, which includes contributions from F- and D-terms, turns
out to be
\beqs \beq \label{Vsusy}V_{\rm SUSY}= \widetilde K^{\al\bbet}
W_\al W^*_\bbet+\frac{g^2}2 \mbox{$\sum_a$} {\rm D}_a {\rm
D}_a\eeq
where $\widetilde K$ is the limit of $K$'s in Eqs.~(\ref{K1}) --
(\ref{K7}) for $\mP\to\infty$ which is
\beq \label{Kquad}\widetilde K=\cm F_- -N\cp F_+ +|S|^2\,.\eeq
Upon substitution of $\widetilde K$ into \Eref{Vsusy} we obtain
\beq V_{\rm SUSY}=\ld^2\left|\phcb\phc-\frac{M^2}4\right|^2
+\frac{\ld^2}{\cm(1-N\rs)}|S|^2\lf|\phcb|^2+|\phc|^2\rg+
\frac{g^2}2\cm^2(1-N\rs)^2\lf|\phcb|^2-|\phc|^2\rg^2\,.
\label{VF}\eeq\eeqs  From the last equation, we find that the SUSY
vacuum lies along the D-flat direction $|\phcb|=|\phc|$ with
\beq \vev{S}\simeq 0 \>\>\>\mbox{and}\>\>\>
|\vev{\Phi}|=|\vev{\bar\Phi}|=\Mpq/2,\label{vevs} \eeq
from which we infer that $\vev{\Phi}$ and $\vev{\bar\Phi}$ break
spontaneously $U(1)_{B-L}$, no only during nMHI but also at the
vacuum of the theory. The contributions from the soft SUSY
breaking terms can be safely neglected since the corresponding
mass scale is much smaller than $M$. They may shift
\cite{dvali,r2}, however, slightly $\vev{S}$ from zero in
\Eref{vevs}.

\section{Inflationary Set-up}\label{hi}

In this section, we outline the salient features of our
inflationary scenario. In \Sref{hi1} we derive the tree-level
inflationary potential and in Sec.~\ref{hi2} we consolidate its
stability and its robustness against one-loop radiative
corrections.

\subsection{Tree-level Inflationary Potential}\label{hi1}

If we express $\Phi, \bar\Phi$ and $S$ according to the
parametrization
\beq\label{hpar} \Phi=\frac{\sg e^{i\th}}{\sqrt{2}}
\cos\thn,~~\bar\Phi=\frac{\sg e^{i\thb}}{\sqrt{2}}
\sin\thn,~~\mbox{and}~~S=\frac{s +i\bar s}{\sqrt{2}}\,,\eeq
with $0\leq\thn\leq\pi/2$, we can easily deduce from \Eref{Vsugra}
that a D-flat direction occurs at
\beq \label{inftr} \bar
s=s=\th=\thb=0\>\>\>\mbox{and}\>\>\>\thn={\pi/4}\eeq
along which the only surviving term in \Eref{Vsugra} can be
written universally as
\beq \label{Vhio}\Vhi=e^{K}K^{SS^*}\,
|W_{,S}|^2\,=\frac{\ld^2(\sg^2-M^2)^2}{16\fr^{2(1+n)}}\,,~~~\mbox{where}~~~
\fr=1+\cp\sg^2\eeq
plays the role of a non-minimal coupling to Ricci scalar in the
\emph{Jordan frame} ({\sf\ftn JF}). Indeed, if we perform a
conformal transformation \cite{linde1, lazarides} defining the
frame function as
\beq\label{omgdef}
{\Omega/N}=-\exp\lf-{K}/{N}\rg,~~\mbox{where}~~N=N_i~~
\mbox{for}~~K=K_i~~\mbox{with}~~i=1,...,7,\eeq
we can easily show that $\fr=-\Omega/N$ along the path in
\Eref{inftr}. Since $M\ll1$, we obtain $\vev{\fr}\simeq1$ at the
SUSY vacuum in \Eref{vevs} and therefore the conventional Einstein
gravity is recovered. For the derivation of \Eref{Vhio}, we also
set
\beq \label{Nab} n=\begin{cases} {(N_i-1)/2}-1\\{N_{2+j}/2}-1
\end{cases} \hspace*{-0.3cm}\mbox{and}~~K^{SS^*}=\begin{cases} \fr\\1\end{cases}
\mbox{for}~~ K=\begin{cases} K_i&\mbox{with}\>\>i=1,2\\
K_{j+2}&\mbox{with}\>\>j= 1,...,5.\end{cases} \eeq Note that the
exponent $n$ defined here has not to be confused with the one used
in \cref{nMkin}.

As deduced from \Eref{Vhio} $\Vhi$ is independent from $\cm$ and
$m$ which dominate, though, the canonical normalization of the
inflaton. To specify it, we note that, for all $K$'s in
Eqs.~(\ref{K1}) -- (\ref{K7}), $K_{\al\bbet}$ along the
configuration in \Eref{inftr} takes the form
\beq \lf K_{\al\bbet}\rg=\diag\lf M_K,K_{SS^*}\rg~~\mbox{with}~~
M_K=\frac{1}{\fr^2}\mtta{\kappa}{\bar\kappa}{\bar\kappa}{\kappa},\label{Sni1}
\eeq
where $\kp=\cm\fr^2-N\cp$, $\bar\kp={N\cp^2\sg^2}$.  Upon
diagonalization of $M_K$ we find its eigenvalues which are
\beq
\label{kpm}\kp_+=\frac{\cm}{\fr^2}\lf\fr^{1+m}+N\rs(\cp\sg^2-1)\rg
\>\>\>\mbox{and}\>\>\> \kp_-=\frac{\cm}{\fr}\lf\fr^m-
{N\rs}\rg,\eeq
where the positivity of $\kp_-$ is assured during and after nMHI
for
\beq \label{rsmin}\rs<\fr^m/N~~\mbox{with}~~\rs=\cp/\cm\,.\eeq
Given that $\fr^m>1$ for $m\geq0$ and $\vev{\fr}\simeq1$,
\Eref{rsmin} implies that the maximal possible $\rs$ is $\rs^{\rm
max}\simeq1/N$. As shown numerically in \Sref{res2}, inflationary
solutions with \Eref{rsmin} fulfilled are attained only for
$m\gtrsim-0.6$.

Inserting \eqs{hpar}{Sni1} in the second term of the
\emph{right-hand side} ({\sf\ftn r.h.s}) of \Eref{Saction1} we
can, then, specify the EF canonically normalized fields, which are
denoted by hat, as follows
\beqs\bea \nonumber K_{\al\bbet}\dot z^\al \dot z^{*\bbet}&=&
{\kp_+\over 2}\lf\dot \sg^2+{1\over2}\sg^2\dot\theta^2_+
\rg+{\kp_-\over 2}\sg^2\lf{1\over2}\dot\theta^2_-
+\dot\theta^2_\Phi \rg+\frac12K_{SS^*}\lf\dot s^2+\dot{\bar
s}^2\rg\\&\simeq&\frac12\lf\dot{\widehat \sg}^2+\dot{\widehat
\th}_+^2+\dot{\widehat \th}_-^2+\dot{\widehat
\th}_\Phi^2+\dot{\what s}^2+{\dot{\what{\bar s}}}^{~2}\rg,
\label{Snik}\eea
where $\th_{\pm}=\lf\bar\th\pm\th\rg/\sqrt{2}$,
$K_{SS^*}=1/K^{SS^*}$ with $K^{SS^*}$ being given in \Eref{Nab}
and the dot denotes derivation w.r.t the cosmic time, $t$. Setting
for later convenience $J=\sqrt{\kp_+}$, we can express the hatted
fields in terms of the initial (unhatted) ones via the relations
\beq \label{VJe} \frac{d\se}{d\sg}=J,\>\>\widehat{\theta}_+
={J\over\sqrt{2}}\sg\theta_+,\>\>\widehat{\theta}_-
=\sqrt{\frac{\kp_-}{2}}\sg\theta_-,\>\>\widehat \theta_\Phi =
\sqrt{\kp_-}\sg\lf\theta_\Phi-\frac{\pi}{4}\rg,(\what s,\what{\bar
s})=\sqrt{K_{SS^*}}(s,\bar s)\,.\eeq\eeqs
As we show below the masses of the scalars besides $\se$ during
nMHI are heavy enough such that the dependence of the hatted
fields on $\sg$ does not influence their dynamics -- see also
\cref{nmH}. Note, in passing, that the spinors $\psi_S$ and
$\psi_{\Phi\pm}$ associated with the superfields $S$ and
$\Phi-\bar\Phi$ are normalized similarly, i.e.,
$\what\psi_{S}=\sqrt{K_{SS^*}}\psi_{S}$ and
$\what\psi_{\Phi\pm}=\sqrt{\kp_\pm}\psi_{\Phi\pm}$ with
$\psi_{\Phi\pm}=(\psi_\Phi\pm\psi_{\bar\Phi})/\sqrt{2}$.

\renewcommand{\arraystretch}{1.4}
\begin{sidewaystable}[h!]
\bec
\begin{tabular}{|c|c|c||c|c|c|c|c|}\hline
{\sc Fields}&{\sc Eigen-} & \multicolumn{6}{c|}{\sc Masses
Squared}\\\cline{3-8}
&{\sc states}&& {$K=K_1$}&{$K=K_2$} &{$K=K_{i+2}$}&$K=K_{i+4}$&$K=K_7$ \\
\hline\hline
1 complex scalar&$\widehat s, \widehat{\bar{s}}$ & $ \widehat m_{
s}^2$&\multicolumn{2}{c|}{$6\lf2\kx\fr-1/N_i\rg\Hhi^2$}&$12\kx\Hhi^2$&\multicolumn{2}{|c|}{$(6/\nb)\Hhi^2$}\\\cline{3-8}
2 real scalars&$\widehat\theta_{+}$&$\widehat m_{\theta+}^2$&
$6(1-1/N_1)\Hhi^2$
&\multicolumn{3}{|c|}{$6\Hhi^2$}&$6(1+1/\nb)\Hhi^2$\\\cline{3-8}
&$\widehat \theta_\Phi$ &$\widehat m_{ \theta_\Phi}^2$&
$M^2_{BL}+6(1-1/N_1)\Hhi^2$&
\multicolumn{3}{|c|}{$M^2_{BL}+6\Hhi^2$}&$M^2_{BL}+6(1+1/\nb)\Hhi^2$\\
\hline
1 gauge  boson& $A_{BL}$ &  $
M_{BL}^2$&\multicolumn{5}{c|}{$g^2\cm\lf\fr^{m-1}-N\rpm
/\fr\rg\sg^2$}\\\hline
$4$ Weyl spinors & $\what \psi_\pm $ & $\what m^2_{ \psi\pm}$ &
\multicolumn{2}{c|}{${6(\cp(N-3)\sg^2-2)^2\Hhi^2}/{\cm\sg^2\fr^{1+m}}$}
&\multicolumn{3}{c|}{${6(\cp(N-2)\sg^2-2)^2\Hhi^2}/{\cm\sg^2\fr^{1+m}}$}\\\cline{3-8}
&$\ldu_{BL}, \widehat\psi_{\Phi-}$&
$M_{BL}^2$&\multicolumn{5}{c|}{$g^2\cm\lf\fr^{m-1}-N\rpm
/\fr\rg\sg^2$}\\
\hline
\end{tabular}\eec
\renewcommand{\arraystretch}{1.}
\hfill \caption[]{\sl\small Mass-squared spectrum for $K=K_i,
K_{i+2}, K_{i+4},$ and $K=K_7$ ($i=1,2$) along the inflationary
trajectory in \Eref{inftr} for $\sg\ll1$. $N$ is defined in
\Eref{omgdef} and $\what \psi_\pm =(\what{\psi}_{\Phi+}\pm
\what{\psi}_{S})/\sqrt{2}$. To avoid very lengthy formulas, we
neglect terms proportional to $M\ll\sg$.}\label{tab1}
\end{sidewaystable}
\clearpage

\subsection{Stability and one-Loop Radiative Corrections}\label{hi2}

We can verify that the inflationary direction in \Eref{inftr} is
stable w.r.t the fluctuations of the non-inflaton fields. To this
end we construct the mass-spectrum of the scalars taking into
account the canonical normalization of the various fields in
\Eref{Snik} -- for details see \cref{lazarides}. In the limit
$\cm\gg\cp$, we find the expressions of the masses squared $\what
m^2_{\chi^\al}$ (with $\chi^\al=\theta_+,\theta_\Phi$ and $S$)
arranged in \Tref{tab1}. These results approach rather well the
quite lengthy, exact expressions taken into account in our
numerical computation. From these findings we can easily confirm
that $\what m^2_{\chi^\al}\gg\Hhi^2=\Vhio/3$ during nMHI provided
that $\kx>0.2$ for $K_i$ with $i=1,...,4$ or $0<\nb<6$ for $K_i$
with $i=5,6$ and $7$. In \Tref{tab1} we display also the masses of
the gauge boson $M_{BL}$, which signals the fact that $\Ggut$ is
broken during nMHI, and the masses of the corresponding fermions.
From our results here we can recover those derived in
\cref{nMHkin} for $K_i$ with $i=1,2$ and $N_i=3$ or $i=3,4$ and
$N_i=2$.

The derived mass spectrum can be employed in order to find the
one-loop radiative corrections, $\dV$ to $\Vhi$. Considering SUGRA
as an effective theory with cutoff scale equal to $\mP$ the
well-known Coleman-Weinberg formula \cite{cw} can be employed
self-consistently taking into account the masses which lie well
below $\mP$, i.e., all the masses arranged in \Tref{tab1} besides
$M_{BL}$ and $\what m_{\th_\Phi}$. Following the approach of
\cref{lazarides} we can verify that our results are immune from
$\dV$, provided that the renormalization-group mass scale
$\Lambda$, is determined by requiring $\dV(\sgx)=0$ or
$\dV(\sgf)=0$. The possible dependence of our results on the
choice of $\Lambda$ can be totally avoided if we confine ourselves
to $\kx\sim(0.5-1.5)$ in $K_i$ with $i=1,...,4$ or $0<\nb<6$ in
$K_i$ with $i=5,6$ and $7$ resulting to
$\Ld\simeq(3-5)\cdot10^{-5}$. Under these circumstances, our
results can be reproduced by using $\Vhi$ in \Eref{Vhio}. We
expect that this conclusion is valid even in cases where $\Phi$
and $\bar\Phi$ are charged under more structured gauge groups than
the one adopted here -- see \Sref{fhim}.

\section{Constraining the Parameters of the Models}\label{fhi}

In this section we outline the predictions of our inflationary
scenaria in Secs.~\ref{res1} and \ref{res2}, testing them against
a number of criteria introduced in Sec.~\ref{const}.

\subsection{Observational \& Theoretical Constraints} \label{const}

Our inflationary settings can be characterized as successful if
they can be compatible with a number of observational and
theoretical requirements which are enumerated in the following --
cf. \cref{review}.

\subsubsection{Inflationary {\rm e}-Foldings.} The number of
e-foldings
\begin{equation}
\label{Nhi}  \Ns=\int_{\se_{\rm f}}^{\se_\star}\, d\se\:
\frac{\Ve_{\rm HI}}{\Ve_{\rm HI,\se}}= \int_{\sgf}^{\sgx}\,
J^2\frac{\Ve_{\rm HI}}{\Ve_{\rm HI,\sg}}d\sg\,
\end{equation}
that the pivot scale $\ks=0.05/{\rm Mpc}$ experiences during HI,
has to be enough to resolve the horizon and flatness problems of
standard big bang cosmology, i.e., \cite{plin,nmi}
\begin{equation}  \label{Ntot}
\Ns\simeq61.3+\ln{\Vhi(\sgx)^{1/2}\over\Vhi(\sgf)^{1/4}}+
\frac{1-3w_{\rm rh}}{12(1+w_{\rm rh})}\lf\ln \frac{\pi^2g_{\rm
rh*}\Trh^4}{30\Vhi(\sgf)}-2\ln\fr(\sgf)\rg+{1\over2}\ln{\fr(\sgx)}-\frac1{12}\ln
g_{\rm rh*},
\end{equation}
where we assumed that nMHI is followed in turn by a oscillatory
phase with mean equation-of-state parameter $w_{\rm rh}$
\cite{lazarides}, radiation and matter domination, $\Trh$ is the
reheat temperature after nMHI, $g_{\rm rh*}$ is the energy-density
effective number of degrees of freedom at temperature $\Trh$ --
for the MSSM spectrum we take $g_{\rm rh*}=228.75$. As in
\cref{lazarides} we set $w_{\rm rh}\simeq1/3$ which corresponds to
a quartic potential \cite{turner} and so, $\Ns$ turns out to be
independent of $\Trh$. In \Eref{Nhi} $\sgx~[\sex]$ is the value of
$\sg~[\se]$ when $\ks$ crosses outside the inflationary horizon,
and $\sgf~[\sef]$ is the value of $\sg~[\se]$ at the end of nMHI,
which can be found, in the slow-roll approximation, from the
condition
\beq{\ftn\sf
max}\{\eph(\se),|\ith(\se)|\}\simeq1,\>\mbox{where}\>\>
\eph=\frac12\left(\frac{\Ve_{\rm HI,\se}}{\Ve_{\rm
HI}}\right)^2\>\>\>\mbox{and}\>\> \ith={\Ve_{\rm
HI,\se\se}\over\Ve_{\rm HI}}\cdot \label{srcon} \eeq

\subsubsection{Normalization of the Power Spectrum.} The amplitude $\As$ of the
power spectrum of the curvature perturbation generated by $\sg$ at
the pivot scale $k_\star$ must to be consistent with
data~\cite{plcp}
\begin{equation}  \label{Prob}
\sqrt{A_{\rm s}}=\: \frac{1}{2\sqrt{3}\, \pi} \; \frac{\Ve_{\rm
HI}(\sex)^{3/2}}{|\Ve_{\rm
HI,\se}(\sex)|}=\frac{|J(\sgx)|}{2\sqrt{3}\, \pi} \;
\frac{\Ve_{\rm HI}(\sgx)^{3/2}}{|\Ve_{\rm
HI,\sg}(\sgx)|}\simeq4.627\cdot 10^{-5},
\end{equation}
where we assume that no other contributions to the observed
curvature perturbation exists.

\subsubsection{Inflationary Observables.} The remaining inflationary observables (the spectral index $\ns$,
its running $\as$, and the tensor-to-scalar ratio $r$) must be in
agreement with the fitting of the \plk, \emph{Baryon Acoustic
Oscillations} ({\sf\ftn BAO}) and \bcp\ data \cite{plin,gwsnew}
with $\Lambda$CDM$+r$ model, i.e.,
\begin{equation}  \label{nswmap}
\mbox{\ftn\sf
(a)}\>\>\ns=0.968\pm0.009~~~\mbox{and}~~~\mbox{\ftn\sf
(b)}\>\>r\leq0.07,
\end{equation}
at 95$\%$ c.l. with $|\as|\ll0.01$. Although compatible with
\sEref{nswmap}{b} the present combined \plk\ and \bcp\ results
\cite{gwsnew} seem to favor $r$'s of order $0.01$ since $r=
0.028^{+0.025}_{-0.025}$ at 68$\%$ c.l. has been reported. These
inflationary observables are estimated through the relations:
\beq\label{ns} \mbox{\ftn\sf (a)}\>\>\> \ns=\: 1-6\eph_\star\ +\
2\ith_\star,\>\>\>\mbox{\ftn\sf (b)}\>\>\>
\as=\:\frac23\left(4\widehat\eta_\star^2-(\ns-1)^2\right)-2\widehat\xi_\star\>\>\>\mbox{and}\>\>\>\mbox{\ftn\sf
(c)}\>\>\>r=16\eph_\star, \eeq
where $\widehat\xi={\Ve_{\rm HI,\widehat\sg} \Ve_{\rm
HI,\widehat\sg\widehat\sg\widehat\sg}/\Ve_{\rm HI}^2}$ and the
variables with subscript $\star$ are evaluated at $\sg=\sgx$. For
a direct comparison of our findings with the obervational outputs
in \cref{plin,gwsnew}, we also compute $\rw=16\eph(\se_{0.002})$
where $\se_{0.002}$ is the value of $\se$ when the scale
$k=0.002/{\rm Mpc}$, which undergoes $\what N_{0.002}=\Ns+3.22$
e-foldings during nMHI, crosses the horizon of nMHI.

\subsubsection{Tuning of the Initial Conditions.} For $n>0$ and $m>0$,
$\Vhi$ develops a local maximum
\beq \Vhi(\sg_{\rm max})=\frac{\ld^2 n^{2 n}}{16 \cp^2(1 + n)^{2
(1 + n)}}~~\mbox{at}~~ \sg_{\rm max}=\frac1{\sqrt{\cp
n}}\,,\label{Vmax}\eeq
giving rise to a stage of hilltop \cite{lofti} nMHI. In a such
case we are forced to assume that \FHI occurs with $\sg$ rolling
from the region of the maximum down to smaller values. Therefore a
mild tuning of the initial conditions is required which can be
quantified somehow defining \cite{gpp} the quantity:
\beq \Dex=\left(\sg_{\rm max} - \sgx\right)/\sg_{\rm
max}\,.\label{dms}\eeq
The naturalness of the attainment of \FHI increases with $\Dex$
and it is maximized when $\sg_{\rm max}\gg\sgx$ which result to
$\Dex\simeq1$.

\subsubsection{Gauge Unification.} To determine better our models
we specify $M$ involved in \Eref{Win} by requiring that
$\vev{\Phi}$ and $\vev{\bar\Phi}$ in \Eref{vevs} take the values
dictated by the unification of the MSSM gauge coupling constants,
despite the fact that $U(1)_{B-L}$ gauge symmetry does not disturb
this unification and $M$ could be much lower. In particular, the
unification scale $\Mgut\simeq2/2.433\times10^{-2}$ can be
identified with $M_{BL}$ -- see \Tref{tab1} -- at the SUSY vacuum,
\Eref{vevs}, i.e.,
\beq \label{Mg} {\sqrt{\cm(\vev{\fr}^m-N\rs)}gM\over
\sqrt{\vev{\fr}}}=\Mgut\>\Rightarrow\>M\simeq{\Mgut}/{g\sqrt{\cm\lf1-{N\rs}\rg}}\eeq
with $g\simeq0.7$ being the value of the GUT gauge coupling and we
take into account that $\vev{\fr}\simeq1$. This determination of
$M$ influences heavily the inflaton mass at the vacuum and induces
an $N$ dependence in the results which concerns though the
post-inflationary epoch. Indeed, the EF (canonically normalized)
inflaton,
\beq\dphi=\vev{J}\dph\>\>\>\mbox{with}\>\>\> \dph=\phi-M
\>\>\>\mbox{and}\>\>\>\vev{J}=\sqrt{\vev{\kp_+}}\simeq\sqrt{1-{N\rs}}\label{dphi}
\eeq
acquires mass, at the SUSY vacuum in \Eref{vevs}, which is given
by
\beq \label{msn} \msn=\left\langle\Ve_{\rm
HI,\se\se}\right\rangle^{1/2}= \left\langle \Ve_{\rm
HI,\sg\sg}/J^2\right\rangle^{1/2}\simeq\frac{\ld
M}{\sqrt{2\cm\lf1-{N\rs}\rg}}\,,\eeq
where the last (approximate) equalities above are valid only for
$\rs\ll1/N$ -- see \eqs{kpm}{VJe}. Upon substitution of the last
expression in \Eref{Mg} into \Eref{msn} we can infer that $\msn$
remains constant for fixed $\rs$ since $\ld/\cm$ is fixed too --
see \Sref{res1}.

\subsubsection{Effective Field Theory.}  To avoid corrections from quantum
gravity and any destabilization of our inflationary scenario due
to higher order non-renormalizable terms -- see \Eref{Win} --, we
impose two additional theoretical constraints on our models --
keeping in mind that $\Vhi(\sg_{\rm f})\leq\Vhi(\sg_\star)$:
\beq \label{subP}\mbox{\ftn\sf (a)}\>\> \Vhi(\sgx)^{1/4}\leq1
\>\>\>\mbox{and}\>\>\>\mbox{\ftn\sf (b)}\>\> \sgx\leq1.\eeq
The \emph{ultaviolet} ({\ftn\sf UV}) cutoff of our model is $1$
(in units of $\mP$) and so no concerns regarding the validity of
the effective theory arise. Indeed, the fact that $\dphi$ in
\Eref{dphi} does not coincide with $\dph$ at the vacuum of the
theory -- contrary to the pure nMHI \cite{cutoff, riotto} --
assures that the corresponding effective theories respect
perturbative unitarity up to $\mP=1$ although $\cm$ may take
relatively large values for $\sg<1$ -- see \Sref{res1}. To clarify
further this point we analyze the small-field behavior of our
models in the EF. Although the expansions presented below, are
valid only during reheating we consider the $\Qef$ extracted this
way as the overall cut-off scale of the theory since reheating is
regarded \cite{riotto} as an unavoidable stage of nMHI. We focus
first on the second term in the r.h.s of \Eref{Saction1} for
$\mu=\nu=0$ and we expand it about $\vev{\phi}=M\ll1$ in terms of
$\se$. Our result can be written as
\beqs\beq\label{Jexp}  J^2
\dot\phi^2\simeq\lf1+(m-1)\rs\what{\sg}^2+3N\rs^2\what{\sg}^2+\lf1-\frac12m(m-3)\rg\rs^2\what{\sg}^2-5N\rs^3\sg^4+\cdots\rg\dot\se^2.\eeq
Expanding similarly $\Vhi$, see \Eref{Vhio}, in terms of $\se$ we
have
\beq
\Vhi\simeq\frac{\ld^2\what{\sg}^4}{16\cm^{2}}\lf1-2(1+n)\rs\what{\sg}^{2}+(3+5n)\rs^2\what{\sg}^4-\cdots\rg\,.
\label{Vexp}\eeq\eeqs
From the expressions above we conclude that our models are
unitarity safe up to $\mP$ for $\rs\leq1$ and $m$ not much larger
than unity.

\subsection{Analytic Results} \label{res1}

Neglecting $M\ll1$ -- determined as shown above -- from the
expression of $\Vhi$ in \Eref{Vhio} and approximating adequately
$J$ in \Eref{VJe} we can obtain an understanding of the
inflationary dynamics which is rather accurate in the cases
studied below. Since positivity of $\kp_-$ in \Eref{kpm} requires
$m\gtrsim-0.6$ -- see \Sref{res2} -- we disregard the tiny allowed
region with $m<0$ from our analytic treatment. In addition, given
that analytic results for $m=0$ and $n\leq0$ are worked out in
\cref{lazarides} we here focus on $m>0$. As for $m=0$, the first
term in the r.h.s of the expression of $\kp_+$ in \Eref{kpm} is by
far the dominant one and so $J$ is well approximated by
\beq J\simeq\sqrt{{\cm}{\fr^{m-1}}}. \label{J1}\eeq
Obviously, $J$ is $n$ independent and for $m=1$ it becomes $\phi$
independent too. Using this estimation, the slow-roll parameters
can be calculated as follows
\beq \label{srg}\eph=\frac{8(1-n\cp\sg^2)^2}{\cm\sg^2 \fr^{1+m}}
\>\>\>\mbox{and}\>\>\> \ith=4\:\frac{3 - (2 + m +9n)\cp\sg^2 +n(m
+ 4 n)\cp^2\sg^4}{\cm\sg^2 \fr^{1+m}}\,\cdot \eeq
Expanding $\eph$ and $\ith$ for $\sg\ll1$ we can find that
\Eref{srcon} entails
\beq \sgf\simeq\mbox{\ftn\sf max}\left\{\frac{2
\sqrt{2/\cm}}{\sqrt{1+ 8(1 + m + 2 n)\rs}},\frac{2
\sqrt{3/\cm}}{\sqrt{1+ 4 (5 + 4 m + 9 n)\rs}}\right\}\,.
\label{sgf}\eeq
Moreover, \Eref{Prob} is written as
\beq \label{Proba}\sqrt{\As}=\frac{
\ld\sqrt{\cm}}{32\sqrt{3}\pi}\frac{\sgx^3
\fr(\sgx)^{(m-2n-1)/2}}{1- n\cp \sgx^2}\,\cdot \eeq
As regards $\Ns$, this can be computed from \Eref{Nhi} as follows
\begin{equation}
\label{Nhi1}
\Ns\simeq\int_{\sgf}^{\sgx}d\sg\:\frac{\cm\sg}{4}\frac{\fr^m}{1 -
\cp n \sg^2}\,\cdot
\end{equation}
A comprehensive result for $\Ns$ can be obtained, if we specify
$n$ and $m$. Therefore, we below -- in Secs.~\ref{res1a} and
\ref{res1b} -- focus on two simple cases where informative and
rather accurate results can be easily achieved.

\subsubsection{The $n=0$ Case.}\label{res1a}

In this case, the integration in \Eref{Nhi} can be readily
realized with result
\begin{equation}
\label{Nhia}
\Ns=\frac{\fr(\sgx)^{1+m}-1}{8\rs(1+m)}~~\mbox{with}~~\fr(\sgx)=1+\cp\sgx^2\,,
\end{equation}
given that $\sgx\gg\sgf$. It is then trivial to solve the equation
above w.r.t $\sgx$ as follows
\begin{equation}
\label{sgxa}\sgx\simeq\sqrt{\frac{\fms-1}{\cp}},~~~\mbox{where}~~~
\fms=\lf1+8(m+1)\rs\Ns\rg^{1/(1+m)}\,.
\end{equation}
Obviously there is a lower bound on $\cm$ for every $\rs$ above
which \sEref{subP}{b} is fulfilled. Indeed, from \Eref{sgxa} we
have
\begin{equation}
\label{cmmin}\sgx<1~~\Rightarrow~~\cm\geq{\lf\fms-1\rg}/{\rs}
\end{equation}
and so, our proposal can be stabilized against corrections from
higher order terms of the form $(\phc\phcb)^l$ with $l\geq2$ in
$W$ -- see \Eref{Win}. From \Eref{Prob} we can also derive a
constraint on $\ld/\cm$, i.e.,
\beq \label{lana} \ld=32\pi\sqrt{3\As}\cm
\rs^{3/2}\fms^{{(1-m)}/{2}}/(\fms-1)^{3/2}\,. \eeq
Upon substitution of \Eref{sgxa} into \Eref{ns} we find
\beqs\bea\label{ns1}&& \ns=1 - 8\rs\frac{
m-1+(m+2)\fms}{(\fms-1)\fms^{1+m}},~~r=\frac{128\rs}{(\fms-1)\fms^{1+m}},
\\ &&\as=\frac{64\rs^2(1+m)(m+2)}{(\fms-1)^2\fms^{4(1+m)}}
\fms^2\lf\fms^{2m}\lf\frac{1-m}{m+2}+\frac{2m-1}{m+1}\fms\rg-\fms^{2(1+m)}\rg.~~~
\label{as1}\eea\eeqs
We can clearly infer that increasing $m$ for fixed $\rs$, both
$\ns$ and $r$ increase. Note that this formulae, based on
\Eref{sgxa}, is valid only for $\rs>0$ (and $m\neq0$). Obviously,
our present results reduce to those displayed in \cref{nMkin}
performing the following replacements (in the notation of that
paper):
\beq \label{nmkin} n=4,~r_{\cal R\rm K}=\rs,~~\mbox{and}~~c_{\rm
K}=\cm\,\eeq
and multiplying by a factor of two the r.h.s of the equation which
yields $\ld$ in terms of $\cm$. E.g., for $m=1$ we obtain
\beq \label{m1}
\ns\simeq1-\frac{3}{2\Ns}-\frac{3}{8(\Ns^3\rs)^{1/2}},\>\as\simeq-\frac{3}{2\Ns^2}-\frac{9}{16(\Ns^5\rs)^{1/2}},~
r\simeq\frac{1}{2\Ns^2\rs}+\frac{2}{(\Ns^3\rs)^{1/2}} \eeq
in accordance with the findings arranged in Table II of
\cref{nMkin}.

\subsubsection{The $n\neq0$ and $m=1$ Case.}\label{res1b}

In this case, the result of the integration in \Eref{Nhi} for any
$n$ is
\begin{equation}
\label{Nhib}  \Ns\simeq -\frac{n \cp \sgx^2 + (1 + n)\ln(1 - n
\cp\sgx^2)}{8 n^2 \rs}\,,\eeq
where we take into account that $\sgx\gg\sgf$. Solving \Eref{Nhib}
w.r.t $\sgx$ we obtain
\begin{equation}
\label{sgxb}\sgx\simeq
{\sqrt{\frs-1\over\cp}}~~\mbox{with}~~\frs={1+n\over
n}\lf1+W_k\lf{y\over1+n}\rg\rg.
\end{equation}
Note that $n \rs\sex^2<1$ for all relevant cases. Here $W_m$ is
the Lambert $W$ or product logarithmic function \cite{wolfram}
with $y=-\exp\lf-(1+8n^2\Ns\rs)/(1+n)\rg$. We take $k=0$ for
$n\geq0$ and $k=-1$ for $n<0$. As in the case above, $\sgx\leq1$
is assured if we impose a lower bound on $\cm$ given by
\Eref{cmmin} replacing $\fms$ with $\fns$.

Upon substitution of \Eref{sgxb} into \Eref{Proba} we obtain a
constraint on $\ld/\cm$, i.e.
\beq \ld=32\sqrt{3\As}\pi
\cm\rs^{3/2}\frs^n\frac{n(1-\frs)+1}{(\frs-1)^{3/2}}\,\cdot
\label{lanb}\eeq
Plugging also \Eref{sgxb} into the definitions of the inflationary
observables -- see \Eref{ns} -- and expanding successively the
exact result for low $n$ and $1/\Ns$ we find
\beqs\bea\nonumber &&\ns=1-8\rs\frac{3\fns-\lf\fns^2+\fns-2\rg n+2n^2\lf\fns-1\rg^2}{\fns^2(\fns-1)} \\
&&~~~~\simeq1-4n^2\rs-2n\frac{\rs^{1/2}}{\Ns^{1/2}}-\frac{3-2n}{2\Ns}-\frac{3-n}{8(\Ns^3\rs)^{1/2}}\,,\label{ns2}\\
&& r=\frac{128\rs\lf1+n(1-\fns)\rg^2}{\fns^2(\fns-1)}\simeq
-\frac{8n}{\Ns}+\frac{3+2n}{6\Ns^2\rs}+\frac{6-n}{3(\Ns^3\rs)^{1/2}}
+\frac{8n^2\rs^{1/2}}{\Ns^{1/2}}\,,\label{rs2}\\
&& \as\simeq
-\frac{n\rs^{1/2}}{\Ns^{3/2}}-\frac{3-2n}{2\Ns^2}\,\cdot\label{as2}\eea\eeqs
For $n=0$, $\fns$ in \Eref{sgxb} and our outputs in
Eqs.~(\ref{ns2}) -- (\ref{as2}) coincide with $f_{m*}$ and the
corresponding findings obtained in \Sref{res1a}. Increasing $m$
above $1$ we expect that we will obtain qualitatively similar
results without their analytic verification to be probably
feasible.

\subsection{Numerical Results}\label{res2}

Adopting the definition of $n$ in \Eref{Nab}, our models, which
are based on $W$ in \Eref{Win} and the $K$'s in \Eref{K1} --
(\ref{K7}), can be universally described by the following
parameters:
\bea\nonumber\ld,\>n,\>m,\>\cm,\>\cp\>\mbox{and}\>\>\kx\>\mbox{or}\>\nb.\eea
for the $K$'s given by Eqs.~(\ref{K1}) -- (\ref{K4}) or
Eqs.~(\ref{K5}) -- (\ref{K7}), respectively. Note that $M$, which
is determined by \Eref{Mg}, does not affect the inflationary
dynamics since $M\ll\phi$ during nMHI. Moreover, $\kx$ or $\nb$
influences only $\what m_{s}^2$ in \Tref{tab1} and lets intact the
inflationary predictions provided that these are selected so that
$\what m_{s}^2>$. Performing, finally, the rescalings
$\Phi\to\Phi/\sqrt{\cm}$ and $\bar\Phi\to\bar\Phi/\sqrt{\cm}$, in
Eqs.~(\ref{Win}) and (\ref{K1}) -- (\ref{K7}) we see that, for
fixed $n$ and $m$, $W$ and the $K$'s depend exclusively on
$\ld/\cm$ and $\rs$ respectively. Under the same condition, $\Vhi$
in \Eref{Vhio} is a function of $\rs$ and $\ld/\cm$ and not $\cm$,
$\cp$ and $\ld$ as naively expected.

In our numerical computation we substitute $\Vhi$ from
Eq.~(\ref{Vhio}) in Eqs.~(\ref{Nhi}), (\ref{srcon}), and
(\ref{Prob}), and we extract the inflationary observables as
functions of $n, \rs$, $\ld/\cm$, and $\sgx$. The two latter
parameters can be determined by enforcing the fulfillment of
Eqs.~(\ref{Ntot}) and (\ref{Prob}). We then compute the
predictions of the model for $\ns$ and $r$ constraining from
Eq.~(\ref{nswmap}) $n$ and $\rs$ for every selected $m$. Moreover,
\sEref{subP}{b} bounds $\cm$ from below, as seen from
\Eref{cmmin}. Finally, \Eref{rsmin} provides an upper bound on
$\rs$, which is slightly $N$ dependent. Just for definiteness we
clarify here that our results correspond to the $K$'s given by
Eqs.~(\ref{K3}) -- (\ref{K7}), unless otherwise stated.

\begin{figure}[!t]\vspace*{-.12in}
\hspace*{-.19in}
\begin{minipage}{8in}
\epsfig{file=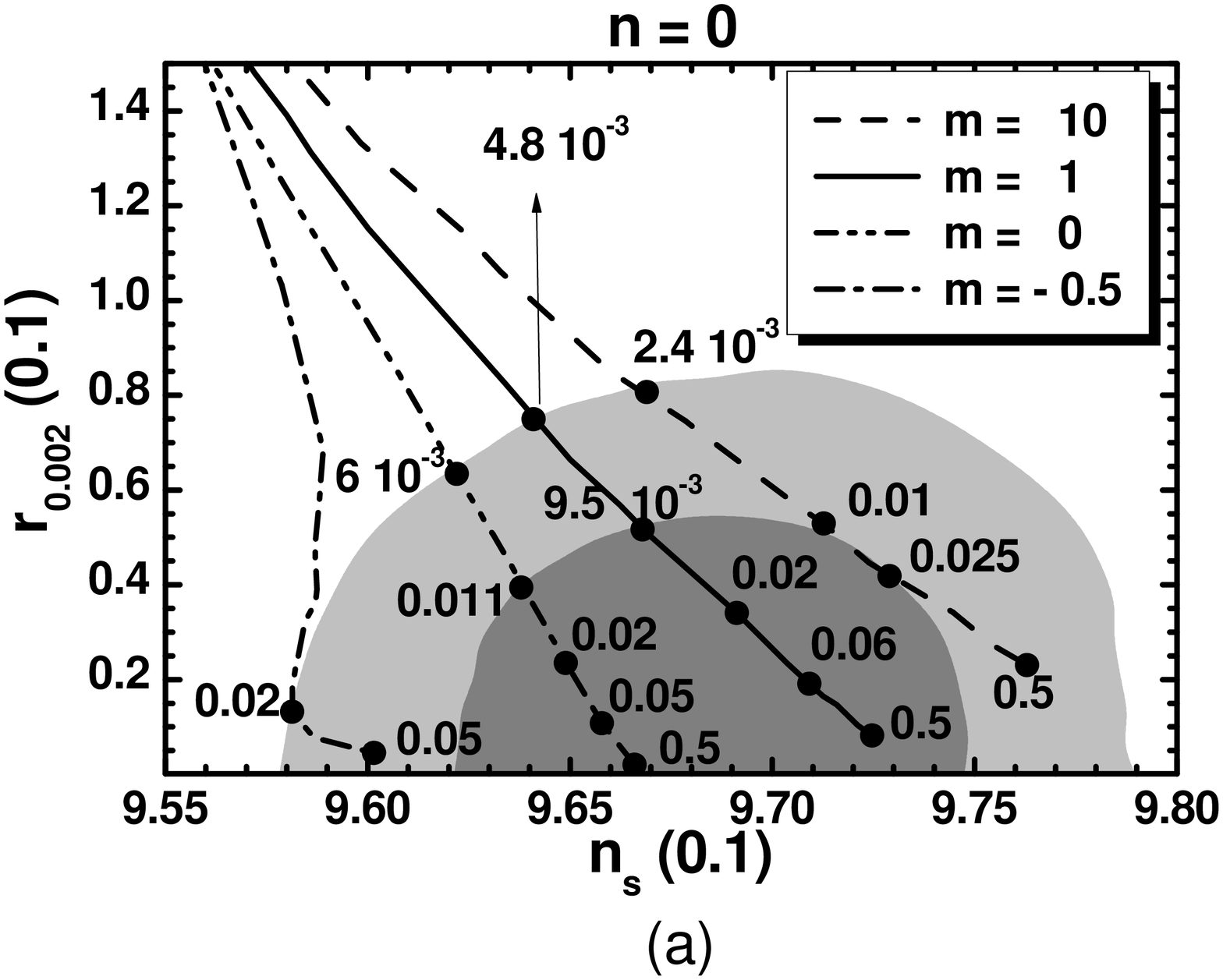,height=3.6in,angle=-90}
\hspace*{-1.2cm}
\epsfig{file=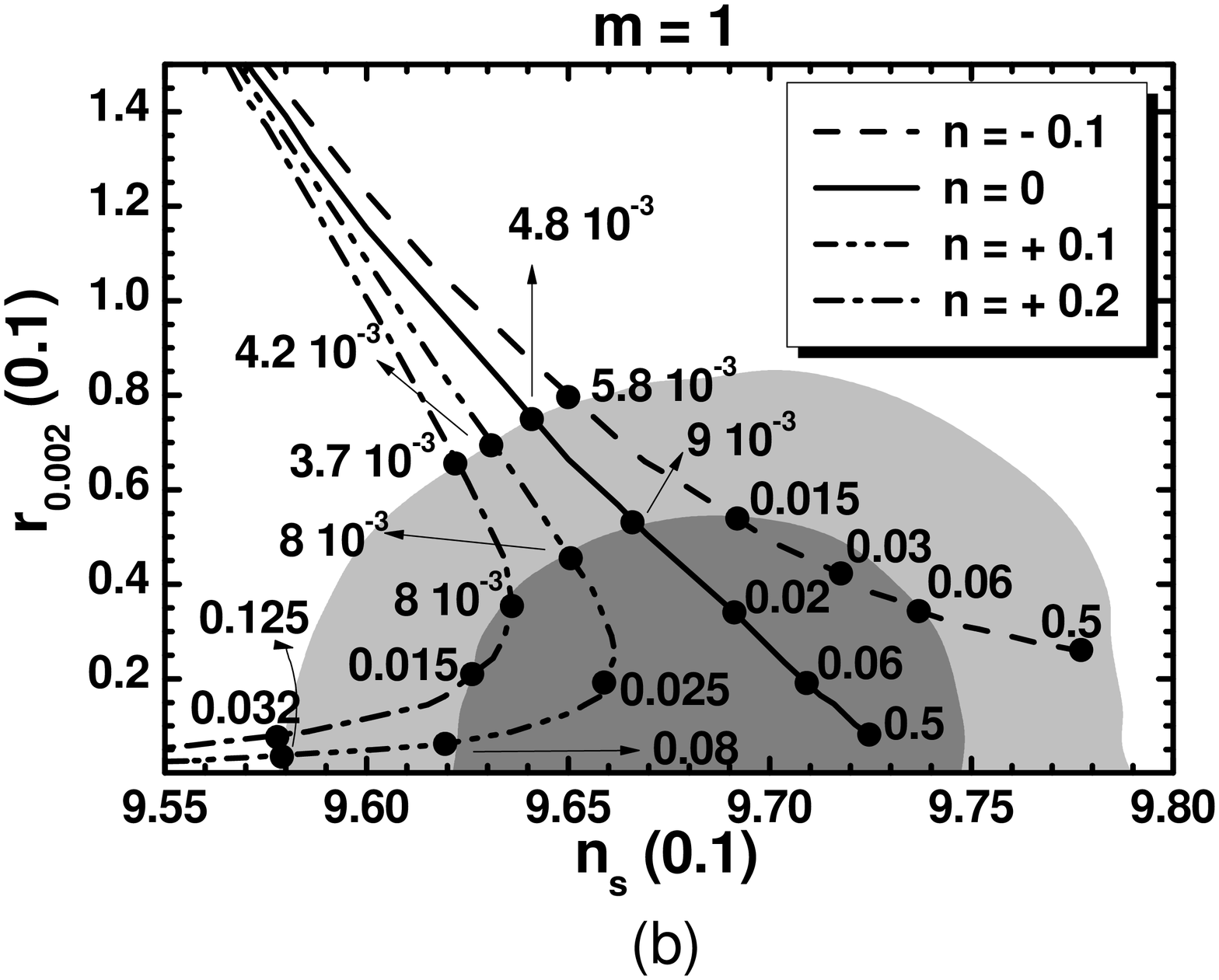,height=3.6in,angle=-90} \hfill
\end{minipage}
\hfill\begin{center}\renewcommand{\arraystretch}{1.1}
\begin{tabular}{|c||cccc|cccc|}\hline
{\sc Plot} &\multicolumn{4}{c|}{{\ftn \sf (a)}: $n=0$ \& $m$ {\sc
Equal to:}}&\multicolumn{4}{c|}{{\ftn \sf (b)}: $m=1$ \& $n$ {\sc
Equal to:}}\\\cline{2-9}
&$-0.5$&$0$&$1$&$10$&$-0.1$&$0$&$0.1$&$0.2$\\\hline\hline
$r_\pm^{\rm min}/10^{-3}$&$20$&$6$&$4.8$&$2.4$&$5.8$&$4.8$&$4.2$&$3.7$\\
$r_\pm^{\rm max}$&$0.05$&$0.5$&$0.5$&$0.5$&$0.5$&$0.5$&$0.125$&$0.032$\\
$\rw^{\rm
min}/10^{-3}$&$3.9$&$1.9$&$6.5$&$23$&$25$&$6.5$&$3.5$&$7.9$
\\\hline
\end{tabular}
\end{center}\vspace*{-.08in}
\hfill \caption{\sl\small Allowed curves in the $\ns-\rw$ plane
for $n=0$ and $m=-1, 0, 1, 10$  {\sffamily\ftn (a)}  or $m=1$ and
$n=-0.1, 0, 0.1, 0.2$ {\sffamily\ftn (b)} with the $\rs$ values
indicated on the curves. The conventions adopted for the various
lines are also shown. The marginalized joint $68\%$ [$95\%$]
regions from \plk, \bcp\ and BAO data are depicted by the dark
[light] shaded contours. The allowed $r_\pm^{\rm min}$ and
$r_\pm^{\rm max}$ together with the minimal $\rw$, $\rw^{\rm
min}$, in each case are listed in the table.}\label{fig1}
\end{figure}\renewcommand{\arraystretch}{1.}


We start the presentation of our results by comparing the outputs
of our models against the observational data \cite{plin,gwsnew} in
the $\ns-\rw$ plane -- see \Fref{fig1}. We depict the
theoretically allowed values with dot-dashed, double dot-dashed,
solid and dashed lines respectively for {\ftn\sf (i)} $n=0$ and
$m=-0.5,0,1$ and $10$ in \sFref{fig1}{a} or {\ftn\sf (ii)} $m=1$
and $n=1/5,1/10,0$ and $-1/10$ in \sFref{fig1}{b}. The variation
of $\rs$ is shown along each line. In both plots, for low enough
$\rs$'s -- i.e. $\rs\leq0.0005$ -- the various lines converge to
$(\ns,\rw)\simeq(0.947,0.28)$ obtained within quatric inflation
defined for $\cp=0$. Increasing $\rs$ the various lines enter the
observationally allowed regions, for $\rs$ equal to a minimal
value $\rs^{\rm min}$, and cover them. The lines corresponding to
$n=0$ and $m=0,1,10$ or $m=1$ and $n=0,-0.1$ terminate for
$\rs=\rs^{\rm max}\simeq0.5$, beyond which \Eref{rsmin} is
violated. The same origin has the termination point of the line
corresponding to $n=0$ and $m=-0.5$ which occurs for $\rs=0.05$.
Finally the lines drawn with $m=1$ and $n=1/5$ or $n=1/10$ cross
outside the allowed corridors and so the $r_{\pm}^{\rm max}$'s,
are found at the intersection points. More specifically, the
values of $\rs^{\rm min}$ and $\rs^{\rm max}$ for any line
depicted in \Fref{fig1}, are accumulated in the Table shown below
the plots -- the entries of the fourth and seventh column coincide
with each other, since in both cases we have $m=1$ and $n=0$.

From \sFref{fig1}{a} we deduce that increasing $m$ above $-0.5$
with $n=0$ the various curves move to the right. On the other
hand, from \sFref{fig1}{b} we infer that for $m=1$ the lines with
$n>0$ [$n<0$] cover the left lower [right upper] corner of the
allowed range. Obviously for $m>1$ we expect that solutions with
$n>0$ are preferable since they fill the observationally favored
region -- cf. \Fref{fig4} below. As we anticipated in
\Sref{const}, for $n>0$ nMHI is of hilltop type. The relevant
parameter $\Dex$ ranges from $0.07$ to $0.66$ for $n=1/10$ and
from $0.19$ to $0.54$ for $n=1/5$ where $\Dex$ increases as $\rs$
drops. That is, the required tuning is not severe mainly for
$\rs<0.1$. In conclusion, the observationally favored region can
be wholly filled varying conveniently $m$ for $n=0$ or $n$ for
$m=1$.

\begin{figure}[!t]\vspace*{-.25in}
\begin{center}
\epsfig{file=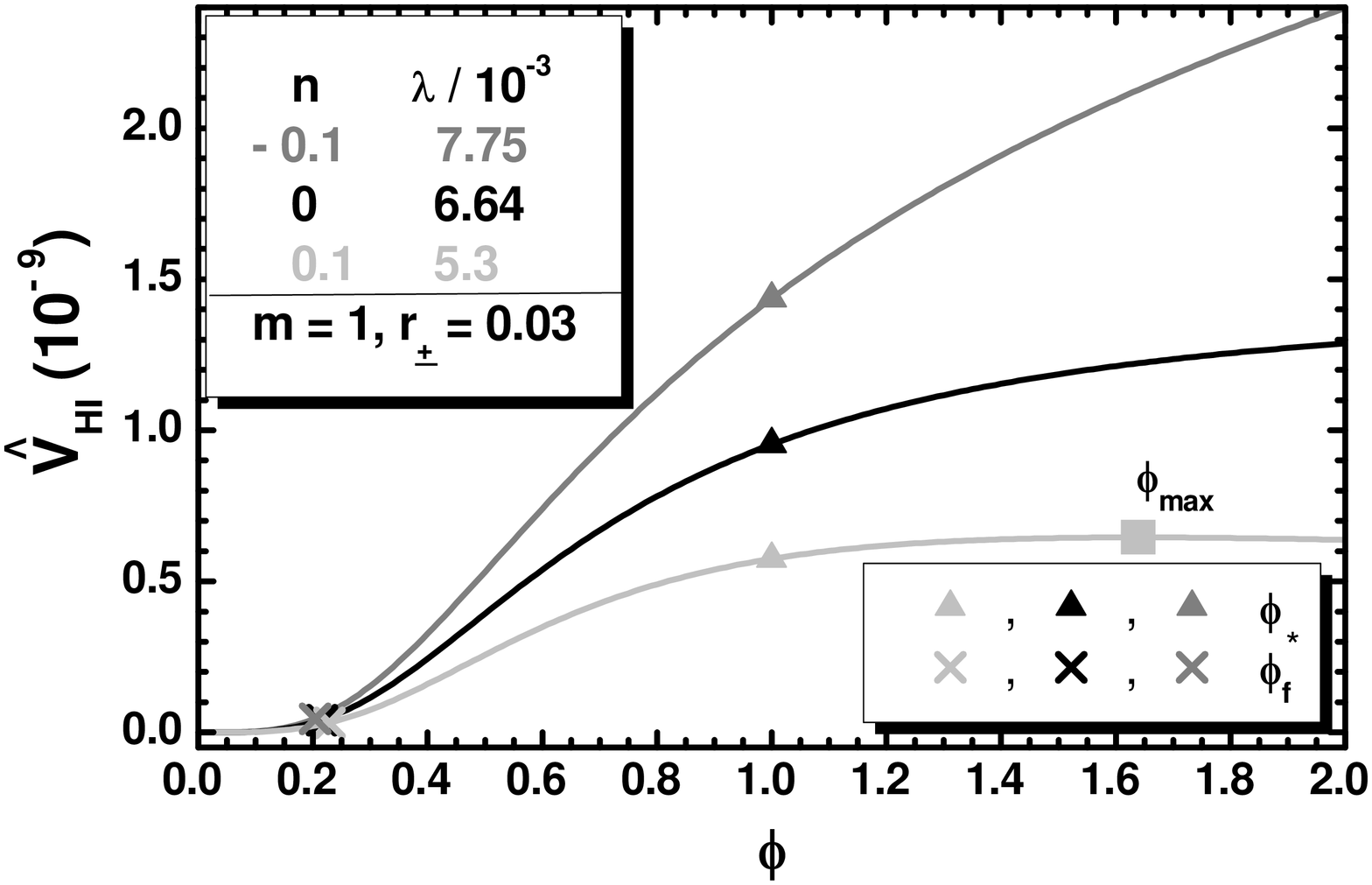,height=3.65in,angle=-90}\eec
\vspace*{-.2in}\hfill  \caption[]{\sl \small The inflationary
potential $\Vhi$ as a function of $\sg$ for $\sg>0$ and  $m=1$,
$\rs\simeq0.03$, and $n=-0.1$, $\ld=7.75\cdot 10^{-3}$ (gray
line), $n=0$, $\ld=6.64\cdot 10^{-3}$ (black line), or $n=+0.1$,
$\ld=5.3\cdot 10^{-3}$ (light gray line). The values of $\sgx$,
$\sgf$ and $\sg_{\rm max}$ (for $n=1/10$) are also
indicated.}\label{fig2}
\end{figure}

The structure of $\Vhi$ as a function of $\sg$ for $\sgx=1$,
$\rs=0.03$, $m=1$ and $n=-0.1$ (light gray line), $n=0$ (black
line) and $n=0.1$ (gray line) is displayed in \Fref{fig2}. The
corresponding values of $\ld$ are $(7.75, 6.64~{\rm or}~5.3)\cdot
10^{-3}$ with $\cm$ being calculated from \Eref{Prob} to be $(1.7,
1.46,{\rm or}~1.24)\cdot 10^2$ whereas the corresponding
observable quantities are found numerically to be $\ns=0.971,
0.969$ or $0.966$ and $r=0.045, 0.03$ or $0.018$ with
$\as\simeq-5\cdot10^{-4}$ in all cases. These results are
consistent with the analytic formulas of \Sref{res1}. Indeed,
applying them we find $\ns=0.97,0.969$ or $0.965$ and $r=0.047,
0.031$ or $0.019$ in excellent agreement with the numerical
outputs above. We observe that $\Vhi$ is a monotonically
increasing function of $\sg$  for $n\leq0$ whereas it develops a
maximum at $\sg_{\rm max}=1.64$, for $n=0.1$, which leads to a
mild tuning of the initial conditions of nMHI since $\Dex=39\%$,
according to the criterion discussed in \Sref{const}. It is also
remarkable that $r$ increases with the inflationary scale,
$\Vhi^{1/4}$, which in all cases approaches the SUSY GUT scale
$M_{\rm GUT}\simeq 8.2\cdot10^{-3}$ as expected -- see e.g.
\cref{rRiotto}.

The relatively high $r$ values encountered here are associated
with \trns\ values of $\sex$ in accordance with the Lyth bound
\cite{lyth}. Indeed, in all cases $\sex\simeq\sqrt{\cm}\sgx>1$ as
can be derived from \Eref{J1}. This fact, though, does not
invalidate our scenario since $\Phi$ and $\bar\Phi$ remain
subplanckian thanks to \sEref{subP}{b} which is satisfied imposing
a lower bound  on $\cm$ -- see e.g. \Eref{cmmin} -- although
$\sex>1$. A second implication of \sEref{subP}{b} is that although
$\ld/\cm$ is constant for fixed $\rs$, $n$ and $m$, the amplitudes
of $\ld$ and $\cm$ can be bounded. E.g., for $n=0, m=1$ and
$\rs=0.03$ we obtain $146\lesssim\cm\lesssim7\cdot10^6$ for
$6.6\cdot10^{-4}\lesssim\ld\lesssim3.5$, where the upper bound
ensures that $\ld$ stays within the perturbative region.

Concentrating on the most promising cases with $n=0$ or $m=1$, we
delineate, in \Fref{fig3}, the allowed regions of our models by
varying continuously $\rs$ and $m$ for $n=0$, in \sFref{fig3}{a},
or $n$ for $m=1$, in \sFref{fig3}{b}. The conventions adopted for
the various lines are also shown in the figure. In particular, the
allowed (shaded) regions are bounded by the dashed line, which
originates from \Eref{rsmin}, and the dot-dashed and thin lines
along which the lower and upper bounds on $\ns$ and $r$ in
\Eref{nswmap} are saturated respectively. We remark that
increasing $\rs$, with $n=0$ and fixed $m$, $r$ decreases, in
accordance with our findings in \sFref{fig1}{a}. On the other
hand, for $m=1$, $\rs$ takes more natural -- in the sense of the
discussion below \Eref{K7} -- values (lower than unity) for larger
values of $|n|$ where hilltop nMHI is activated. Fixing $\ns$ to
its central value in \Eref{nswmap} we obtain the thick solid lines
along which we get clear predictions for $m$ in \sFref{fig3}{a} or
$n$ in \sFref{fig3}{b}, $\rs$ and the remaining inflationary
observables. Namely, from \sFref{fig3}{a}, for $n=0$ and
$\Ns\simeq58$, we obtain
\beqs\beq\label{resn0}  0.2\lesssim m\lesssim4,\>\>\>
0.064\lesssim {\rs\over0.1}\lesssim5,\>\>\> 0.29\lesssim
{r\over0.01}\lesssim7\>\>\>\mbox{and}\>\>\>0.28\lesssim
10^5{\ld\over \cm}\lesssim1.9\,. \eeq
Comparing \sFref{fig3}{a} with Fig. 2 of \cref{nMHkin} we see that
the latest \cite{gwsnew} upper bound on $r$ in \Eref{nswmap} cuts
the lower right slice from the allowed region and consequently a
part from the solid line. Also the allowed region is limited to
$m\gtrsim-0.6$ since below this value \Eref{rsmin} is broken, as
we now recognize. Similarly, from \sFref{fig3}{b}, for
$\ns=0.968$, $m=1$ and $\Ns\simeq58$ we find
\beq\label{resm1} -1.21\lesssim {n\over0.1}\lesssim0.215,\>\>\>
0.12\lesssim {\rs\over0.1}\lesssim5,\>\>\> 0.4\lesssim
{r\over0.01}\lesssim7\>\>\>\mbox{and}\>\>\>0.25\lesssim
10^5{\ld\over \cm}\lesssim2.6\,. \eeq\eeqs Hilltop nMHI is
attained for $0<n\leq0.0215$ and there, we get $\Dex\gtrsim0.4$.
In both cases above $\as$ is confined in the range
$-(5-6)\cdot10^{-4}$ and so, our models are consistent with the
fitting of data with the $\Lambda$CDM+$r$ model \cite{plin}.
Moreover, our models are testable by the forthcoming experiments
\cite{ictp} searching for primordial gravity waves since
$r\gtrsim0.0019$.

\begin{figure}[!t]\vspace*{-.12in}
\hspace*{-.19in}
\begin{minipage}{8in}
\epsfig{file=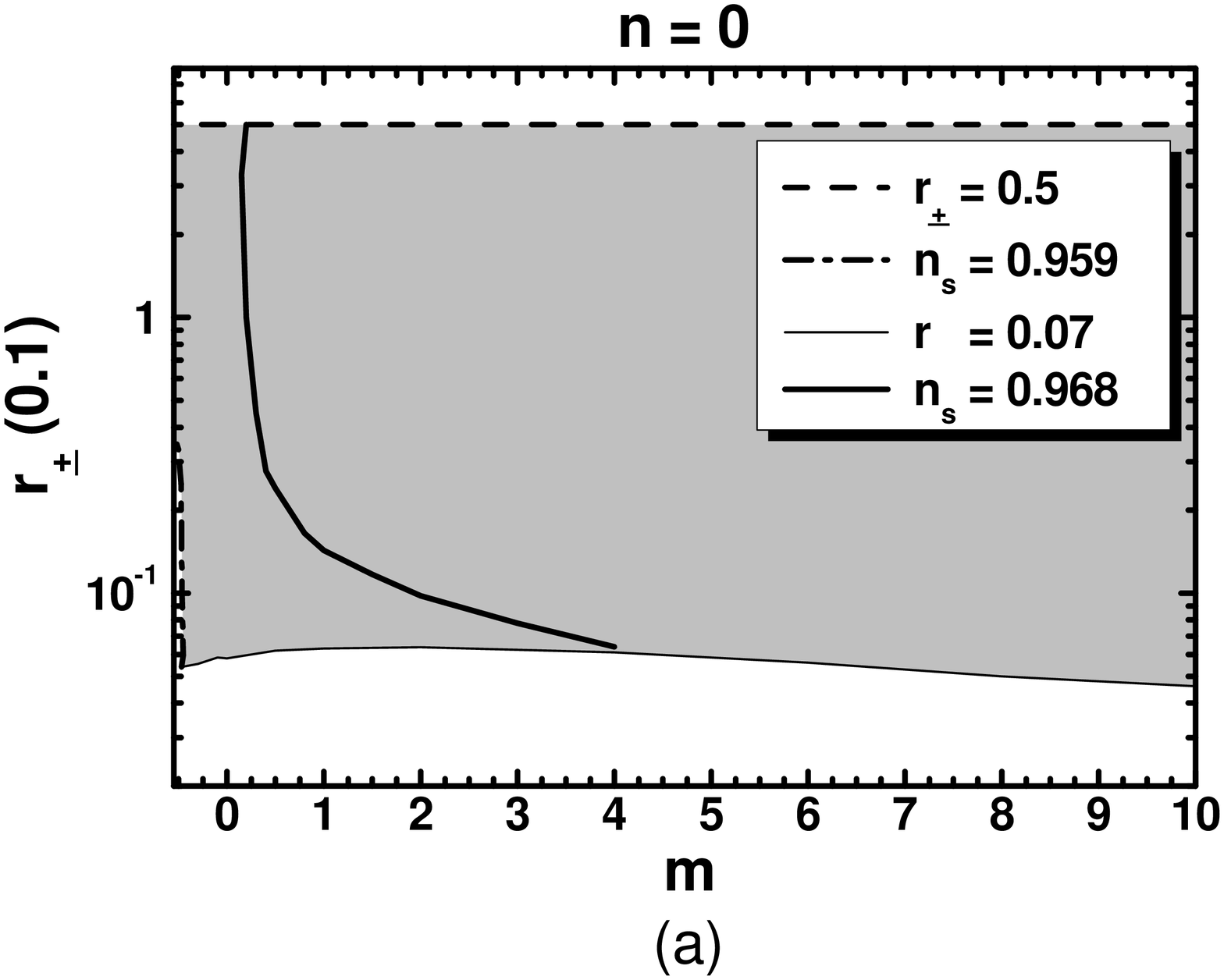,height=3.6in,angle=-90}
\hspace*{-1.2cm}
\epsfig{file=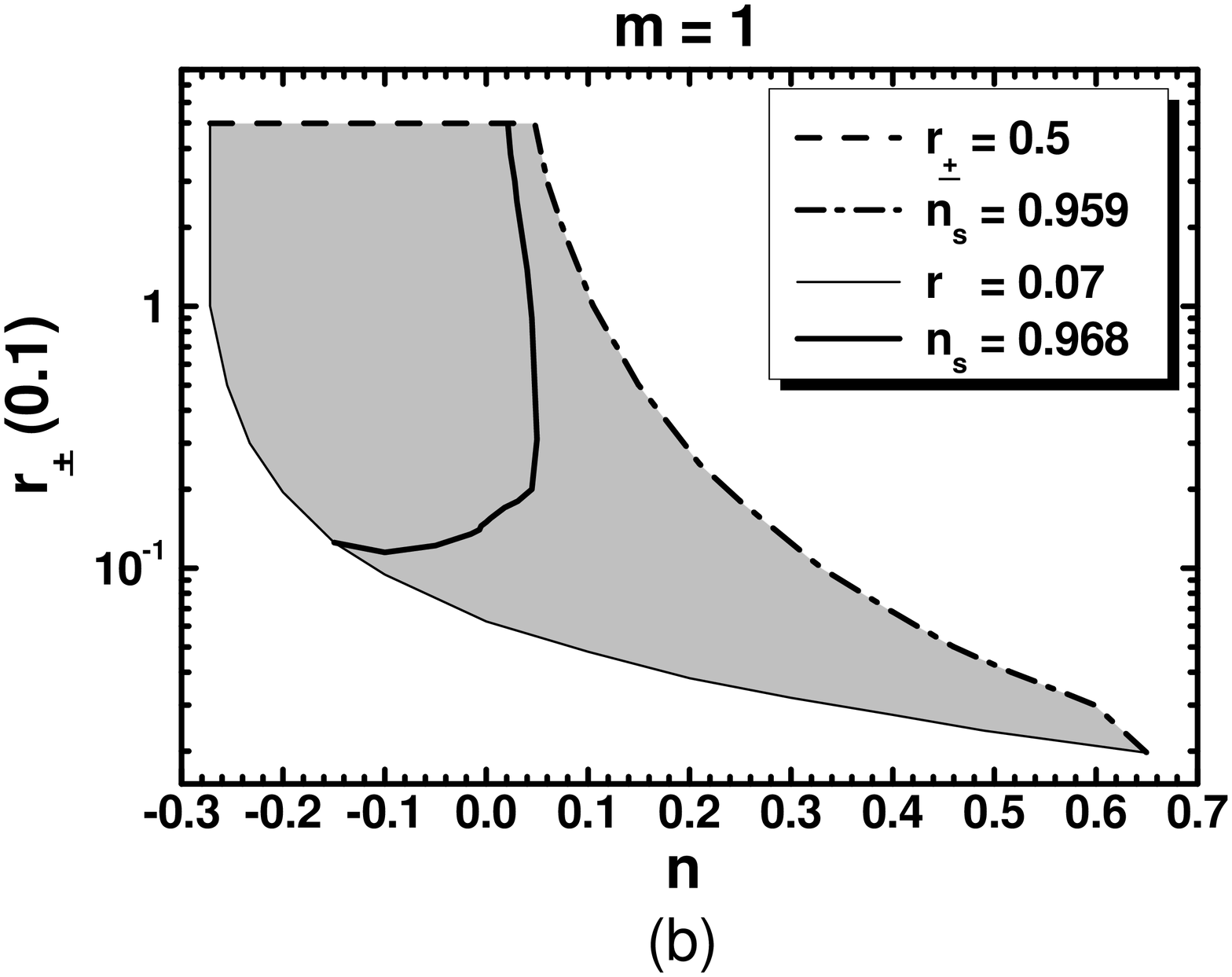,height=3.6in,angle=-90} \hfill
\end{minipage}
\hfill \caption{\sl\small Allowed (shaded) regions in the $n-\rs$
plane for $n=0$ {\sffamily\ftn (a)} and $m=1$ {\sffamily\ftn (b)}.
The conventions adopted for the various lines are also
shown.}\label{fig3}
\end{figure}

Had we employed $K_i$ with $i=1,2$, the various lines ended at
$\rs\simeq0.5$ in \Fref{fig1} and the allowed regions in
Fig.~\ref{fig3} would have been shortened until $\rs\simeq1/3$.
This bound would have yielded slightly larger $\rw^{\rm min}$'s.
Namely, $\rw^{\rm min}/10^{-3}\simeq2.8, 8.4$ and $25$ for $n=0$
and $m=0,1$ and $10$ whereas $\rw^{\rm min}\simeq0.026$ for $m=1$
and $n=-0.1$ -- the $\rw^{\rm min}$'s for $n>0$ are let
unaffected. For $n=0$ and $m=-0.5$ we obtain $\rs^{\rm
max}\simeq0.04$ and $\rw^{\rm min}\simeq0.0045$. The lower bound
of $r/0.01$ and the upper ones on $\rs/0.1$ and $10^5\ld/\cm$ in
\Eref{resn0} [\Eref{resm1}] become $0.42$, $3.3$ and $1.5$
[$0.64$, $3.3$ and $2.1$] whereas the bounds on $\as$ remain
unaltered.

Fixing $\rs$ to some representative value, we can delineate the
allowed region of our models in the $m-n$ plane as shown in
\Fref{fig4}. Namely we set $\rs=0.008$ in \sFref{fig4}{a} and
$\rs=0.03$ in \sFref{fig4}{b}. We use the same shape code for the
the boundary lines of the allowed (shaded) regions as in
Fig.~\ref{fig3}. Particularly, the dot-dashed thick line
corresponds to the lower bound on $\ns$ in \sEref{nswmap}{a}
whereas the thin line comes from \sEref{nswmap}{b}. Along the
solid thick line the central value of $\ns$ in \sEref{nswmap}{a}
is attained. We see that the largest parts of the allowed regions
are found for $n>0$ which means that nMHI is of hilltop type.
Moreover, comparing \sFref{fig4}{a} and \sFref{fig4}{a} we remark
that the $n>0$ slice of the allowed region is extended as $\rs$
decreases. In all, for $\ns=0.968$ we take:
\beqs\bea\label{resmn1} && 2.3\lesssim {r/
0.01}\lesssim7\>\>\>\mbox{with}\>\>\>-0.08\lesssim
n\lesssim1.69\>\>\>\mbox{and}\>\>\> 2\lesssim
{m}\lesssim10\, \>\>\>(\rs=0.008);\\
\label{resmn2} && 1.2\lesssim {r/
0.01}\lesssim2.2\>\>\>\mbox{with}\>\>\>-0.0135\lesssim
n\lesssim1.46\>\>\>\mbox{and}\>\>\> 0\lesssim {m}\lesssim10\,
\>\>\>(\rs=0.03). \eea\eeqs
From the relevant plots we observe that $n$ increases with $m$
along the bold solid line. Hilltop nMHI is attained for $m\geq3$
with $\Dex\gtrsim0.26$ for $\rs=0.008$ and for $m\geq0.45$ with
$\Dex\gtrsim0.58$ for $\rs=0.03$.  In both cases, $\Dex$ (and $r$)
decreases as $m$ increases.

As we mention in \Sref{const}, $\msn$ is affected heavily from the
choice of $K$'s in Eqs.~(\ref{K1}) -- (\ref{K7}) as $\rs$
approaches its upper bound in \Eref{rsmin}. Particularly, if we
employ $K_i$ with $i=3,...,7$ along the bold solid lines in
\sFref{fig3}{a} and \sFref{fig3}{b} we obtain
\beqs\beq\label{resmass} 2.4\cdot10^{-3}\lesssim {\msn/
10^{-5}}\lesssim1.2\,\>\>\>\mbox{and}\>\>\>2.1\cdot10^{-3}\lesssim
{\msn/10^{-5}}\lesssim1.5\, \eeq
respectively, whereas for $i=1,2$ the upper bounds above remain
unchanged and the lower bounds move on to $2.6\cdot10^{-2}$ and
$2.4\cdot10^{-2}$ correspondingly. On the other hand, along the
bold solid lines in \sFref{fig4}{a} and \sFref{fig4}{b} we obtain
\beq\label{resmassnm} 2\lesssim {\msn/
10^{-8}}\lesssim4.9\,\>\>\>\mbox{and}\>\>\>2.9\lesssim
{\msn/10^{-8}}\lesssim12\, \eeq\eeqs
respectively, with the bounds being independent from the choice of
$K$. These $\msn$ ranges let open the possibility of non-thermal
leptogenesis \cite{ntlepto} if we introduce a suitable coupling
between $\bar\Phi$ and the right-handed neutrinos -- see e.g.
\crefs{fhi1,nmH}.

\begin{figure}[!t]\vspace*{-.12in}
\hspace*{-.19in}
\begin{minipage}{8in}
\epsfig{file=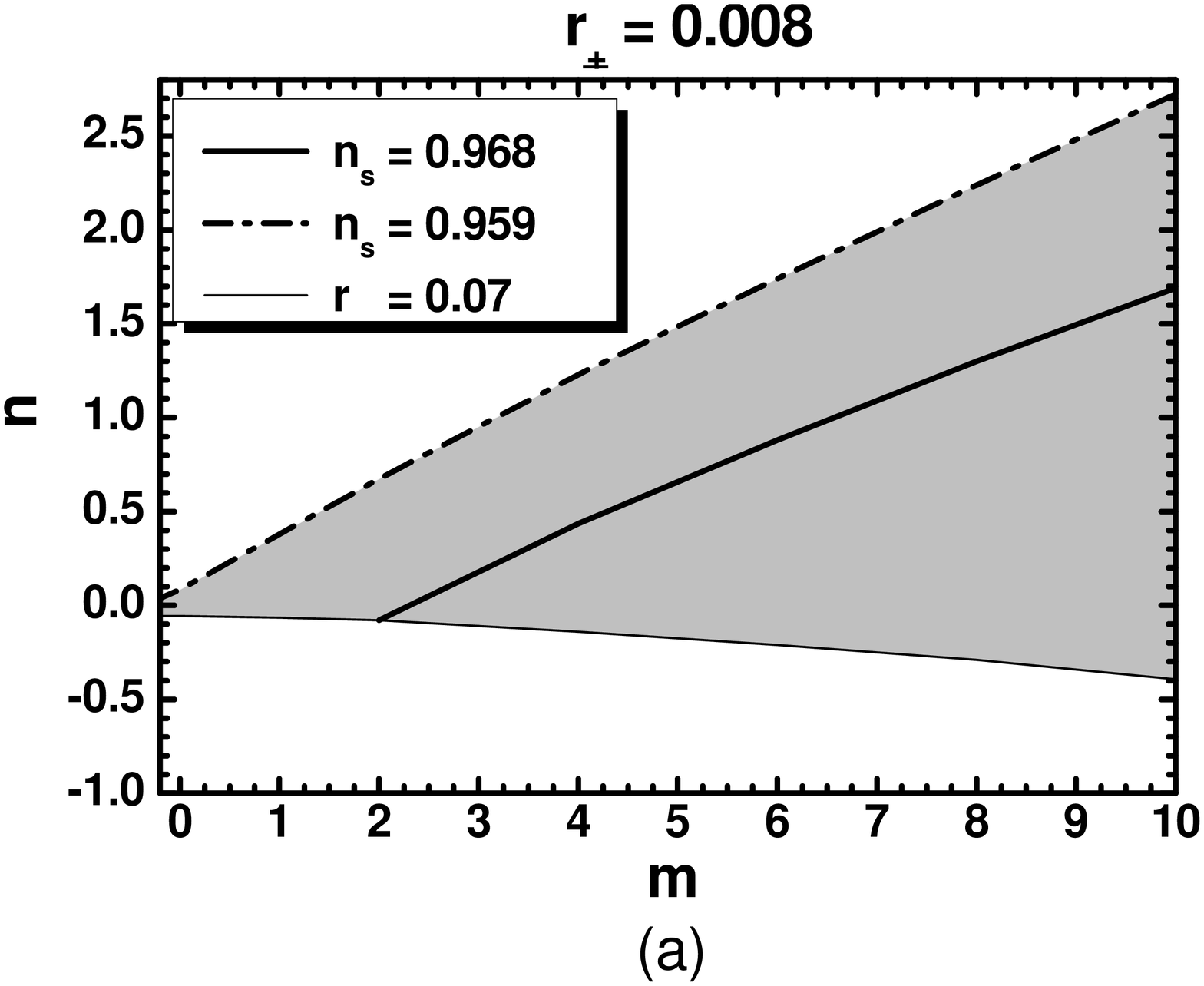,height=3.6in,angle=-90}
\hspace*{-1.2cm}
\epsfig{file=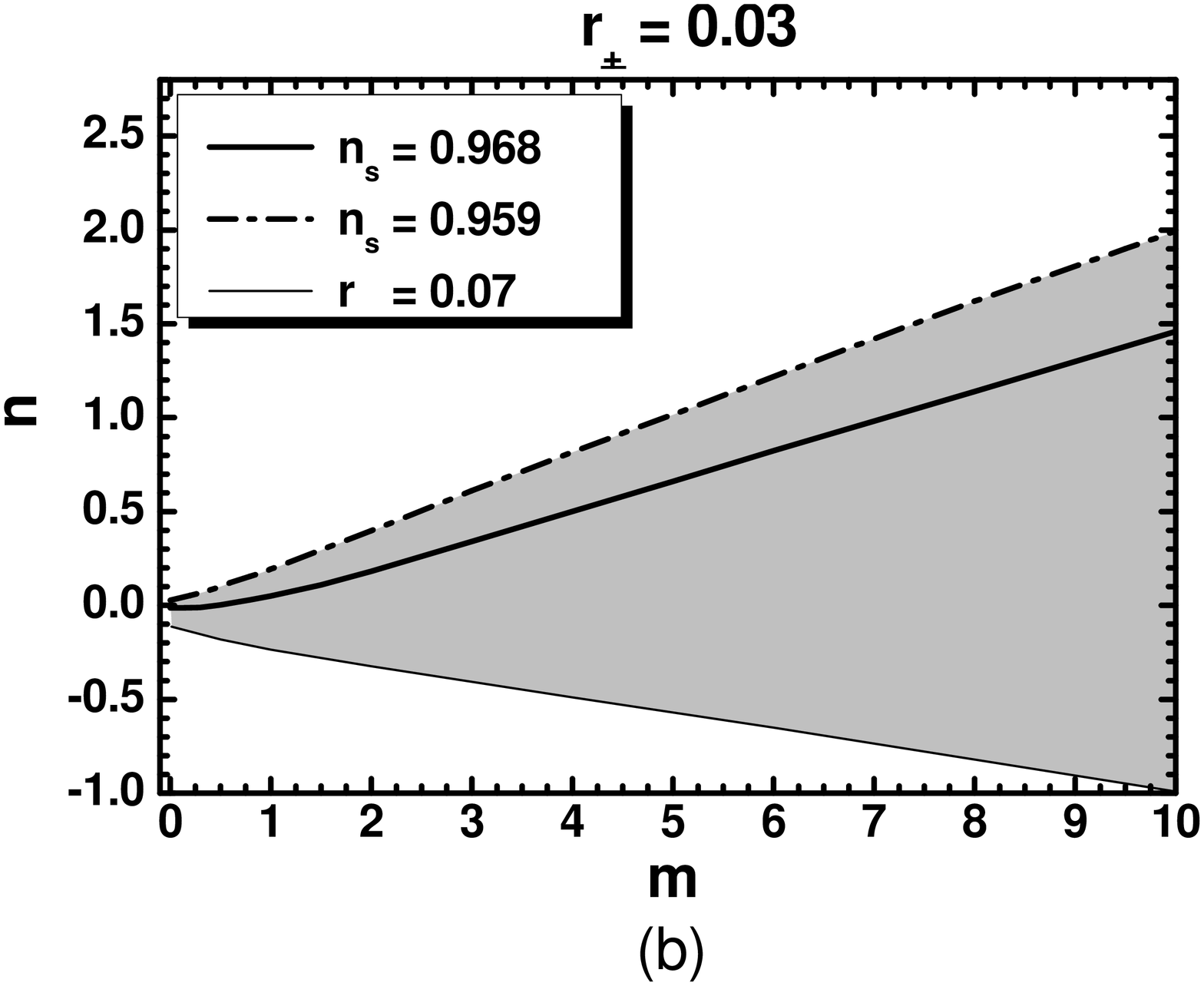,height=3.6in,angle=-90} \hfill
\end{minipage}
\hfill \caption{\sl\small Allowed (shaded) region in the $m-n$
plane for $\rs=0.008$ {\sffamily\ftn (a)} and $\rs=0.03$
{\sffamily\ftn (b)}. The conventions adopted for the various lines
are also shown.}\label{fig4}
\end{figure}

Setting $m=1$ in Eqs.~(\ref{K2}), (\ref{K4}), (\ref{K6}) or
(\ref{K7}) and $n=0$ -- i.e. $N_2=3$ in Eq.~(\ref{K2}) or $N_i=2$
with $i=4,6,7$ in Eqs.~(\ref{K4}), (\ref{K6}) and (\ref{K7}) -- we
can construct the most economical and predictive version of our
models which evades higher order terms of the form
$(1+\cp\fp)^{m-1}$ and the relevant tuning on $n$. In this
restrictive case, $\ns=0.968$ -- see \Eref{nswmap} -- entails
$\rs=0.015$ and corresponds to $r=0.043$ which is a little higher
than the central observational value -- see details below
\Eref{nswmap} -- but still within the $65\%$ c.l favored margin
\cite{gwsnew}. Moreover, \Eref{rsmin} implies $\rw^{\rm
min}\simeq0.0065$ -- see \Fref{fig1}. The alternative minimalistic
choice $m=n=0$ which avoids higher order terms in Eqs.~(\ref{K1}),
(\ref{K3}) and (\ref{K5}) do not yield solutions with $\ns=0.968$
-- see \sFref{fig3}{a}.

\newpage

\section{Conclusions} \label{con}

Extending our work in \crefs{nMkin,lazarides,nMHkin} we analyzed
further the implementation of kinetically modified nMHI within
SUGRA. We specified seven \Kap s $K_i$ with $i=1,...,7$, see
Eqs.~(\ref{K1}) -- (\ref{K7}), which cooperate with the well-known
simplest superpotential $W$ in \Eref{Win} leading to $\Vhi$,
collectively given in \Eref{Vhio}, and a GUT phase transition at
the SUSY vacuum in \Eref{vevs}. Prominent in the proposed $K$'s is
the role of a shift-symmetric quadratic function $\fm$ in
\Eref{fs} which remains invisible in $\Vhi$ while dominates the
canonical normalization of the Higgs-inflaton. On the other hand,
we employ two stabilization mechanisms for the non-inflaton field
$S$, one with higher order terms, in Eqs.~(\ref{K1}) --
(\ref{K4}), and one leading to a $SU(2)_S/U(1)$ symmetric \Km\ in
Eqs.~(\ref{K5}) -- (\ref{K7}). In all, our inflationary setting
depends essentially on four free parameters ($n$, $m$, $\ld/\cm$
and $\rs$), where $n$ and $\rs$ are defined in terms of the
initial variables as shown in \eqs{Nab}{rsmin} respectively. The
model parameters are constrained to natural values, imposing a
number of observational and theoretical restrictions. Predictions
on $r$ value, testable in the near future, were also obtained.

More specifically, for $n=0$ we updated the results of
\cref{nMHkin} in \sFref{fig1}{a} and \sFref{fig3}{a}. For $n\neq0$
and $m=1$, we found new allowed regions presented in
\sFref{fig1}{b} and \sFref{fig3}{b}. Especially for $n>0$, we
showed that $\Vhi$ develops a maximum which does not disturb,
though, the implementation of hilltop nMHI since the relevant
tuning is mostly very low. Indicatively, fixing $\ns\simeq0.968$
and $n=0$, or $m=1$, or $\rs=0.008$, or $\rs=0.03$ we obtained the
outputs in \Eref{resn0} or \Eref{resm1} or \Eref{resmn1} or
\Eref{resmn2} respectively. The majority of these solutions can be
classified in the hilltop branch as shown in \Fref{fig4} where we
varied continuously $n$ and $m$ with fixed $\rs$.

In all cases, $\ld/\cm$ is computed enforcing \Eref{Prob} and
$|\as|$ turns out to be negligibly small. Our inflationary setting
can be attained with subplanckian values of the initial
(non-canonically normalized) inflaton, requiring large $\cm$'s,
without causing any problem with the perturbative unitarity. It is
gratifying, finally, that our proposal remains intact from
radiative corrections, the Higgs-inflaton may assume ultimately
the v.e.v predicted by the gauge unification within MSSM, and the
inflationary dynamics can be studied analytically and rather
accurately for $n=0$ and $m\geq0$ or $m=1$ and any $n$.

Finally, we would like to point out that, although we have
restricted our discussion on the $\Ggut=G_{\rm SM}\times
U(1)_{B-L}$ gauge group, kinetically modified nMHI analyzed in
this paper has a much wider applicability. It can be realized
within other GUTs, provided that $\Phi$ and $\bar \Phi$ consist a
conjugate pair of Higgs superfields. If we adopt another GUT gauge
group, the inflationary predictions are expected to be quite
similar to the ones discussed here with possibly different
analysis of the stability of the inflationary trajectory, since
different Higgs superfield representations may be involved in
implementing the $\Ggut$ breaking to $G_{\rm SM}$. Removing the
scale $M$ from $W$ in \Eref{Win} and abandoning the idea of grand
unification, our inflationary stage can be realized even by the
electroweak higgs boson -- cf. \cref{shiftHI}. Since our main aim
here is the observational investigation of the kinetically
modified nMHI, we opted to utilize the simplest GUT embedding.

\acknowledgments{The author would like to acknowledge useful
discussions with I. Florakis, D. L\"ust, H. Partouche and
N.~Toumbas and the CERN Theory Division for kind hospitality
during which parts of this work were completed. }


%

\end{document}